\documentclass[aps,pof,reprint,notitlepage,onecolumn,superscriptaddress]{revtex4-1}
\pdfoutput=1
\usepackage{amsmath,graphicx,amssymb}
\usepackage{multirow,color,float}
\usepackage{url}
\usepackage{verbatim}
\usepackage{bm}
\usepackage{subcaption}
\usepackage[normalem]{ulem}

\newcommand{\veps}{\varepsilon}

\newcommand{\etal}{\textit{et al.}}
\renewcommand{\vec}[1]{\bm{#1}}
\newcommand{\fvec}[1]{\hat{\bm{#1}}}

\newcommand{\vu}{\vec{u}}

\newcommand{\vb}{\vec{b}}

\newcommand{\va}{\vec{a}}
\newcommand{\vk}{\vec{k}}
\newcommand{\vp}{\vec{p}}
\newcommand{\vq}{\vec{q}}

\newcommand{\vf}{\vec{f}}
\newcommand{\vw}{\boldsymbol{\omega}}
\newcommand{\avg}[1]{\langle{#1}\rangle}
\newcommand{\dt}{\partial_t}
\def\ukp{u^+_{\bm k}}
\def\ukm{u^-_{\bm k}}

\def\bkp{b^+_{\bm k}}
\def\bkm{b^-_{\bm k}}

\def\usk{u^{s_k}_{\bm k}}

\def\bski{b^{\sigma_k}_{\bm k}}
\def\usp{u^{s_p}_{\bm p}}

\def\bspi{b^{\sigma_p}_{\bm p}}
\def\usq{u^{s_q}_{\bm q}}

\def\bsqi{b^{\sigma_q}_{\bm q}}
\def\hkm{{\bm h}^-_{\bm k}}
\def\hkp{{\bm h}^+_{\bm k}}

\def\hsk{{\bm h}^{s_k}_{\bm k}}
\def\hsp{{\bm h}^{s_p}_{\bm p}}
\def\hsq{{\bm h}^{s_q}_{\bm q}}
\def\hski{{\bm h}^{\sigma_k}_{\bm k}}
\def\Bpp{B^+_{{\bm p}_0}}
\def\Upp{U^+_{{\bm p}_0}}

\def\Bpm{B^-_{{\bm p}_0}}
\def\Upm{U^-_{{\bm p}_0}}

\newcommand{\tuqp}{u^+_{\bm q}}
\newcommand{\tuqm}{u^-_{\bm q}}
\newcommand{\tbkp}{b^+_{\bm k}}
\newcommand{\tbkm}{b^-_{\bm k}}

\newcommand{\beq}{\begin{equation}}
\newcommand{\eeq}{\end{equation}}
\begin{document}
\vspace{-3em}
\begin{minipage}[l]{\textwidth}
Postprint version of the manuscript published in Phys. Rev. Fluids {\bf 2}, 054605 (2017) \\
\end{minipage}
\vspace{1em}
\title{Triad interactions and the bidirectional turbulent cascade of magnetic helicity}

\author{Moritz Linkmann}
\email[]{linkmann@roma2.infn.it}
\affiliation{Department of Physics and INFN, University of Rome Tor Vergata, Via della Ricerca Scientifica 1, 00133 Rome, Italy}
%\affiliation{SUPA, School of Physics and Astronomy, University of Edinburgh, Peter Guthrie Tait Road, EH9 3FD, UK}
\author{Vassilios Dallas}
\email[]{v.dallas@leeds.ac.uk}
\affiliation{Department of Applied Mathematics, University of Leeds, Leeds LS2 9JT, United Kingdom}

\begin{abstract}
Using direct numerical simulations we demonstrate that magnetic helicity
exhibits a bidirectional turbulent cascade at high but finite magnetic
Reynolds numbers.  Despite the injection of positive magnetic helicity in the
flow, we observe that magnetic helicity of opposite signs is generated
between large and small scales.  We explain these observations by carrying out
an analysis of the magnetohydrodynamic equations reduced to %single triads 
triad interactions using the Fourier
helical decomposition.  Within this framework, the direct cascade of positive
magnetic helicity arises through triad interactions that are associated with
small scale dynamo action, while the occurrence of negative magnetic helicity
at large scales is explained through triad interactions that are related to
stretch-twist-fold dynamics and small scale dynamo action, which compete with
the inverse cascade of positive magnetic helicity.  {Our} analytical and
numerical results suggest that the direct cascade of magnetic helicity is a
finite magnetic Reynolds number $Rm$ effect that will vanish in the limit $Rm
\to \infty$.
\end{abstract}

\maketitle

%\noindent Keywords: MHD turbulence, magnetic helicity, turbulent cascades, helical decomposition, triadic interactions

%
\section{Introduction} \label{sec:intro}

The existence of planetary and stellar magnetic fields is currently attributed
to dynamo action \cite{moffatt78,proctorgilbert94}. One of the theoretical
arguments to explain the generation and preservation of magnetic fields in
spatial scales much larger than the outer scales of planets and stars is the
inverse cascade of magnetic helicity in magnetohydrodynamic (MHD) turbulent
flows \cite{Frischetal75}. Magnetic helicity is defined as the correlation
between the magnetic field $\bm b \equiv \nabla \times \bm a$ and the magnetic
potential $\bm a$, i.e., $H_b \equiv \avg{\bm a \cdot \bm b}$, with the angular
brackets denoting a spatial average unless indicated otherwise.
Magnetic helicity is considered to play a critical role in the long-term
evolution of stellar and planetary magnetic fields \cite{brandenburg09} and
hence it is important to understand its dynamics across scales in order to shed
light on the saturation mechanisms of the large-scale magnetic fields of
planets and stars. 

Early studies using turbulent closure models \cite{Pouquet76} and mean field
theory \cite{Steenbeck66,moffatt78,Krause80} have shown (within the framework
of their approximations) that magnetic helicity cascades from small scales to
large scales in agreement with the prediction from equilibrium statistical
mechanics for the ideal MHD equations \cite{Frischetal75}. In these cases, the
inverse cascade of magnetic helicity was associated with the $\alpha$ effect
\cite{parker55} of large scale dynamos, where the kinetic helicity {$H_u =
\avg{\bm u \cdot \bm \omega}$ (where $\bm u$ denotes the velocity field and
$\bm \omega = \nabla \times \bm u$ the vorticity field) generates opposite
signs of magnetic helicity between large and small scales.} This change in sign
across scales is further supported by the magnetic helicity spectra of solar
wind data \cite{Brandenburgetal11}, which show $H_b < 0$ at small wavenumbers
and $H_b > 0$ at large wavenumbers. One way to understand this change in sign
across scales is the conceptual stretch-twist-fold (STF) mechanism
\cite{Childress95,Vainshtein72}. This mechanism proposes that the advection of
magnetic field lines by a positive helical flow leads to a positive magnetic
helicity at small scales and to a negative magnetic helicity at large scales
since large scale magnetic field lines are twisted in the opposite direction.
However, it is presently not clear if such a mechanism is directly associated
with a non-linear cascade process \cite{Mininni11}.

The inverse cascade of magnetic helicity has been verified by direct numerical
simulations (DNSs) \cite{meneguzzietal81,Brandenburg01,Alexakis06} at moderate
values of magnetic Reynolds number $Rm$, where the small-scale magnetic
helicity is dissipated quite fast.  In this case, in contrast with the theory,
a direct cascade of magnetic helicity with smaller magnitude was also observed.
This {bidirectional cascade} of $H_b$, where an inverse and a direct
cascade of magnetic helicity coexist, was observed recently even at high $Rm$
flows \cite{Mueller12,ld16}.  In the limit of high $Rm$ concerns have been
raised about the effectiveness of the $\alpha$ effect in generating large-scale
magnetic fields with strong amplitude due to the detrimental feedback that
fast-growing small-scale magnetic fields have on the growth rate of the large
scales \cite{Vainshtein92,Cattaneo96}. These concerns are supported in some
sense by the bidirectional cascade of magnetic helicity, because the magnitude
of the inverse cascade is limited by the existence of the residual direct
cascade, and by the nonlocality of the inverse cascade of magnetic helicity in
the statistically stationary regime \cite{Alexakis06,ld16}. 

In this paper, we focus on the bidirectional cascade of magnetic helicity and
on the mechanism of magnetic helicity to generate opposite signs of helicity
across the scales. By means of DNSs we inject mean magnetic helicity in our
flows using a helical electromagnetic forcing, while the velocity field is
forced by a nonhelical forcing. This is in contrast to dynamo studies, which
typically force only the velocity field using a helical forcing. In dynamo
studies, the kinetic helicity generates opposite signs of magnetic helicity
between large and small scales.  This sign change across scales makes the
interpretation of statistics ambiguous. With our approach we attempt to have a
dominant sign of magnetic helicity across the scales in order to avoid such
ambiguities as much as possible. The interpretation of our numerical results is
supported by analytical work on the triadic interactions of helical modes in
MHD turbulence \cite{Linkmannetal16,Linkmann17b}. This analysis is based on the
helical decomposition of Fourier modes \cite{Craya58,Lesieur72,Herring74},
which has been an important tool to understand the cascade dynamics of helical
flows in Navier-Stokes turbulence
\cite{Constantin88,Waleffe92,Chen03,Biferale12,Sahoo15,Alexakis17}. Here, the
analysis of triadic interactions of helical modes demonstrates that it can also
be a useful tool to understand further 
(a) the presence of the direct cascade of magnetic helicity to small scales and 
(b) the mechanism that generates opposite signs of magnetic helicity across the scales.

This paper is organized as follows.  We begin with the description of the
numerical method and the resulting database in Sec.~\ref{sec:numerics}. 
Sections ~\ref{sec:time_evolution} and \ref{sec:spectral_dynamics} present the global and spectral dynamics from our numerical simulations, respectively.
Our numerical results are interpreted in Sec.~\ref{sec:triadic_analysis} in terms of the triadic interactions of helical modes, where we propose
an explanation for the direct cascade of magnetic helicity and for the mechanism that generates opposite signs of magnetic helicity across the scales. We 
conclude in Sec.~\ref{sec:conclusions} by summarizing our findings. 

\section{Numerical simulations} \label{sec:numerics}
The present work focuses on the 
bidirectional cascade of magnetic helicity. Thus, we consider numerical simulation of three-dimensional MHD turbulent flows forced at intermediate wavenumbers in the absence of large scales condensates.
Forcing at intermediate scales and aiming for a turbulent flow with high enough scale separation poses serious computational constraints, since a very large range of scales needs to be resolved. So, to circumvent the demanding scale separation requirement and to avoid the condensation of energy at large scales, we consider high-order dissipation terms acting at the small and the large scales, respectively. These dissipation terms effectively increase the extent of the inertial range simulated for a given resolution by reducing the range of scales over which dissipation is effective. Therefore, we numerically solve the equations 
\begin{align}
(\partial_t - \nu^- (-1)^{m+1}\Delta^{-m} - \nu^+ (-1)^{n+1}\Delta^{n})\vec{u}&
= \vu \times \vw + \vec{j} \times \vec{b} - \nabla P +\vec{f}_u \nonumber \\
(\partial_t - \eta^- (-1)^{m+1}\Delta^{-m} - \eta^+ (-1)^{n+1}\Delta^{n})\vec{b}&
=  \nabla \times (\vu \times \vb) \hspace{37pt} + \vec{f}_b
\label{eq:mhd}
\end{align}
where $\vec{u}$ denotes the velocity field, $\vec{b}$ the magnetic induction
expressed in Alfv\'{e}n units, $\bm \omega = \nabla \times \bm u$ the
vorticity, $\bm j = \nabla \times \bm b$ the current density, and $P = p +
|\vu|^2/2$, with $p$ the pressure and $\vec{f}_u$ and $\vec{f}_b$ the external
mechanical and electromagnetic forces, respectively.  Energy is dissipated at
the small scales by the terms proportional to $\nu^+$ and $\eta^+$ and at the
large scales by $\nu^-$ and $\eta^-$.  The indices $n$ and $m$ specify the
order of the Laplacian used. In order to obtain a large inertial range, we
choose $n = m = 4$. In the absence of forcing and dissipation Eqs.
\eqref{eq:mhd} reduce to the ideal MHD equations, which conserve the total
energy $E = E_u + E_b = \frac{1}{2} \avg{|\bm u|^2 + |\bm b|^2}$, the magnetic
helicity $H_b = \avg{\bm a \cdot \bm b}$ and the cross-helicity $H_c = \avg{\bm
u \cdot \bm b}$. These conserved quantities are those that determine the
turbulent cascades in our flows.

Equations \eqref{eq:mhd} are solved numerically in a cubic periodic domain with sides of length $2\pi L$ using the standard pseudospectral method, which ensures that $\nabla \cdot \vec{u} = 0$ and $\nabla \cdot \vec{b} = 0$. Full dealiasing is achieved by the $2/3$ rule and as a result the minimum and maximum wavenumbers are $k_{box}$ = 1 and $k_{max}$
= N/3, respectively, where $N$ is the number of grid points in each Cartesian coordinate. Further details of the code can be found in Refs.~\cite{Yoffe12,BereraLinkmann14}.

As it was mentioned above, in these simulations we choose to inject mean magnetic helicity in our flows, attempting to have a dominant sign of magnetic helicity across the scales in order to be able to interpret our results unambiguously in comparison to dynamo studies. 
In order to do this in a systematic way we force $\bm b$ using a helical forcing and $\bm u$ using a non-helical forcing. Here, we choose to force both $\bm u$ and $\bm b$ with the same forcing amplitude, i.e., $|\vf_u| = |\vf_b|$, so that both quantities are dynamically important and $\bm b$ has a nonlinear feedback on the flow through the Lorenz force. The forces $\vf_u$ and $\vf_b$ are constructed from a randomized superposition of eigenfunctions of the curl operator \cite{Brandenburg01,Mueller12,MalapakaMuller13}, resulting in Gaussian distributed and $\delta$ correlated in time forces whose helicities $\langle \vf_{u,b} \cdot \nabla \times \vf_{u,b} \rangle$ and correlation $\langle \vf_u \cdot \vf_b \rangle$ can be exactly controlled. The specific random nature of the forces ensures that at steady state the total energy input rate $\veps = \veps_u + \veps_b = \langle \vu \cdot \vf_u\rangle + \langle \vb \cdot \vf_b\rangle \propto |\vf_u|^2 + |\vf_b|^2$ is known \emph{a priori} \cite{Novikov65}. Therefore, $\veps$ can be used as a control parameter since the amplitudes of the external forces are input parameters. The external forces are chosen such that the net cross helicity in the flow is negligible by keeping the correlation $\avg{\vf_u \cdot \vf_b} = 0$. In summary, no $H_c$ and no $H_u$ are injected into the flow, while the injection of $H_b$ is maximal. Here, we choose to exclude the injection of cross-helicity in the flow because we want to focus on the cascade dynamics of magnetic helicity, which can be influenced by introducing correlations between the velocity and the magnetic field \cite{Frischetal75,Linkmannetal16}. Finally, the initial magnetic and velocity fields are in equipartition with energy spectra peaked at the forcing wavenumber $k_f$ and zero helicities, i.e., $H_b = H_c = H_u = 0$. 

For all simulations, we set the magnetic Prandtl number for the small scales $Pm^+ = \eta^+/\nu^+$ and for the large scales $Pm^-= \eta^-/\nu^-$ to unity, viz. $Pm^+ = Pm^- = 1$. The values of $\nu^-$ and $\eta^-$ are tuned such that the inverse cascade is damped before the largest scales of the system are excited while the values of $\nu^+$ and $\eta^+$ are such that $k_{max}/k_d \geq 1.25$ is satisfied for all the simulations, where $k_d \equiv (\veps/(\nu^+)^3)^{1/(6n-2)}$ is the dissipation wavenumber \cite{BorueOrszag95}.
The magnetic Reynolds number is defined based on the control parameters of the problem as $Rm_f \equiv U k_f^{1-2n}/\eta^+$ with $U \equiv \left(\veps/ k_f \right)^{1/3}$. 

The energy input and thus the dissipation rate $\veps$ is adjusted such that $U$ 
is kept constant while the scale separation $k_fL$ increases, resulting in a set of simulations with the same Reynolds number.
For simulations at $k_fL = 10$ with different Reynolds numbers we kept 
the energy input constant and we varied $\nu^+ = \eta^+$.
All the necessary numerical parameters that make
these simulations reproducible are given in Table~\ref{tbl:simulations} along
with their total runtime $T$ of the simulations normalized by $\tau_f \equiv
(Uk_{box})^{-1}$, a time scale defined based on the control parameters.
%
%%%%%%%%%%%%%%%%%%%%%%%%%%%%%%%%%%%%%%%%%%%%%%%%%%%%%%%%%
 \begin{table}[!htbp]
 \begin{center}
 \begin{tabular}{*{8}{c}}
  $k_fL$ & $N$ & $Rm_f$ & $\nu^+ = \eta^+$ & $\nu^- = \eta^-$ & $|\vf_u|$ & $|\vf_b|$ & $T/t_f$ \\
  \hline
   10 & 128 & $7\times 10^3$ & $6.30 \times 10^{-12}$ & 0.05 & 1 & 1 & 320 \\ %T=750
   20 & 256 & $7\times 10^3$ & $4.92 \times 10^{-14}$ & 0.05 & $\sqrt{2}$ & $\sqrt{2}$ & 130 \\ %T=300 
   40 & 512 & $7\times 10^3$ & $3.68 \times 10^{-16}$ & 0.05 & 2 & 2 &  100 \\ %T=230
  \hline
   10 & 128 & $7\times 10^2$ & $6.30 \times 10^{-11}$ & 0.05 & 1 & 1 & 90 \\ %T=200
   10 & 256 & $9\times 10^4$ & $4.92 \times 10^{-14}$ & 0.05 & 1 & 1 & 110 \\ %T=250 
  \hline
  \end{tabular}
  \end{center}
\caption{
Numerical parameters of the simulations. Note that $k_f$ denotes the forcing
wave number, $T$ is the total runtime in simulation units, $|\vf_u|$ and 
$|\vf_b|$ are 
the mechanical and electromagnetic forcing magnitudes, respectively, and $t_f
\equiv (Uk_{box})^{-1}$ a time scale defined based on the control parameters.
All simulations are well resolved with $k_{cut}/k_d \geqslant 1.25$.  The
hyperdissipative Reynolds number is defined as $Rm_f = U k_f^{1-2n}/\eta^+ $.
}
 \label{tbl:simulations}
 \end{table}
 %%%%%%%%%%%%%%%%%%%%%%%%%%%%%%%%%%%%%%%%%%%%%%%%%%%%%%%%%

%
\section{Time evolution} \label{sec:time_evolution}
\begin{figure}[!ht]
\begin{subfigure}{0.49\textwidth}
  \includegraphics[width=\textwidth]{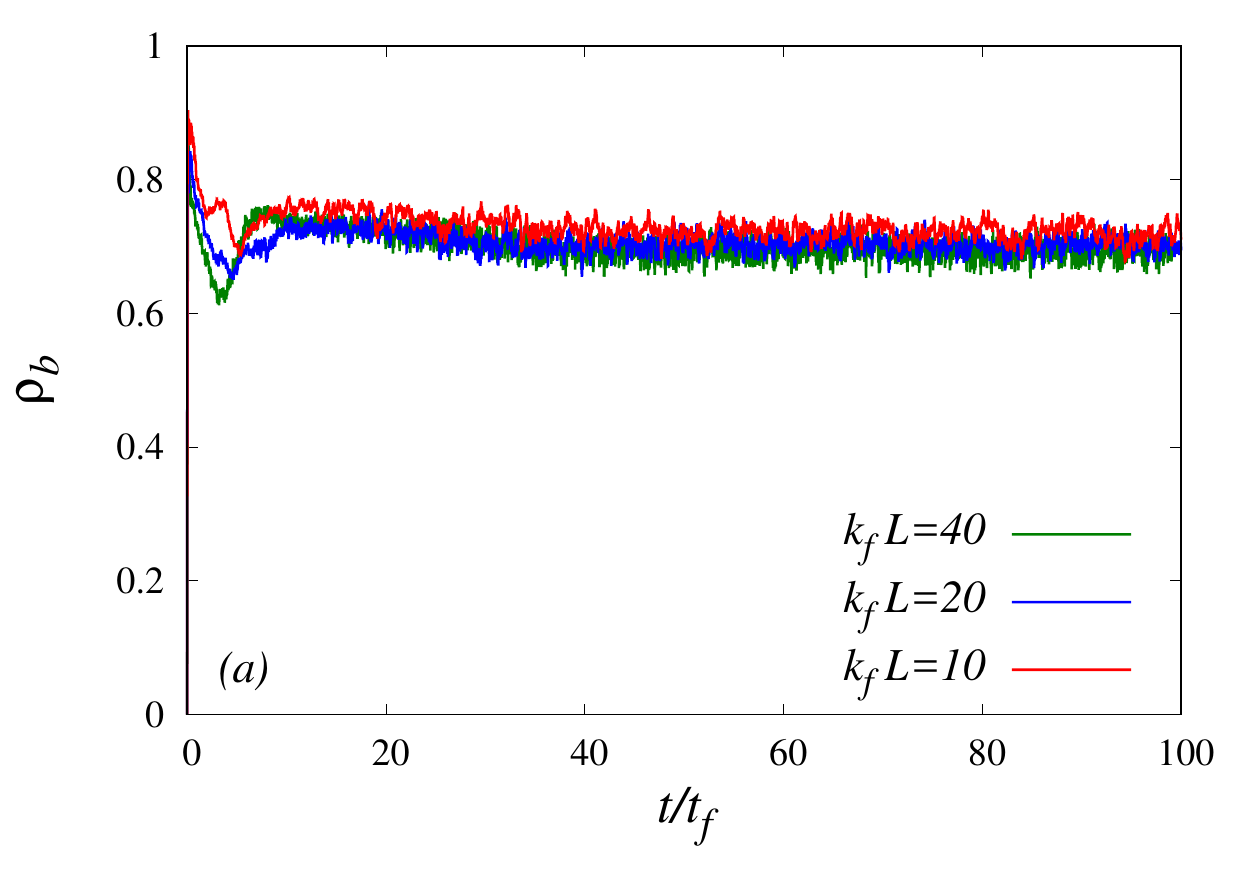}
 \caption{}
 \label{fig:rhob}
\end{subfigure}
\begin{subfigure}{0.49\textwidth}
  \includegraphics[width=\textwidth]{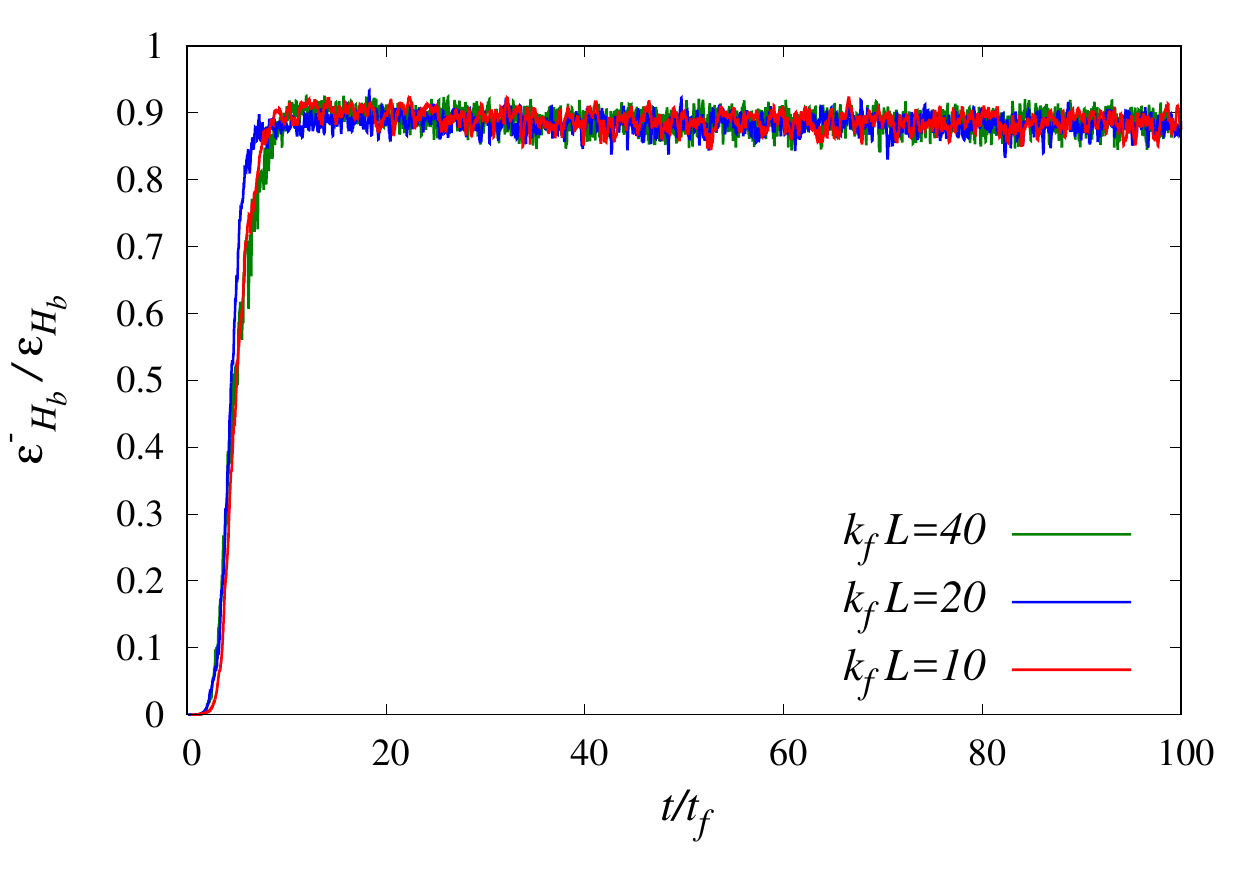}
 \caption{}
 \label{fig:epshb}
\end{subfigure}
 \caption{Time series of 
(a) the normalized magnetic helicity $\rho_b$ and 
(b) the ratio $\veps_{H_b}^-/\veps_{H_b}$ for flows with scale separations $k_fL = 10, 20$ and 40.}
 \label{fig:timeseries}
\end{figure}

Before discussing the interaction of the helical modes in order to understand further the dynamics of magnetic helicity across scales, we provide an overview of the statistics of the flows under study. 
In Fig. \ref{fig:rhob} we have plotted the time evolution of the normalized magnetic helicity
\begin{equation}
 \rho_b \equiv H_b / (\avg{|\bm a|^2}\avg{|\bm b|^2})^{1/2},
\end{equation}
which belongs to the range $-1 \leq \rho_b \leq 1$. For $\rho_b = 0$ the flows have no magnetic helicity, while for $\rho_b = \pm 1$ the flows are fully dominated by magnetic helicity, which means that the Lorentz force $\bm j \times \bm b$ is expected to be zero on average.
At time $t=0$, we set $\rho_b = 0$ but instantaneously it reaches high values because of the direct injection of positive $H_b$ by the helical electromagnetic force.
At steady state $\rho_b$ reaches a mean value of 0.7 for all the flows with scale separations $k_fL = 10, 20, 40$, indicating that the flows are dominated to a large degree by positive mean magnetic helicity. %Thus, now we can discuss unambiguously the cascade dynamics of $H_b$.

The strength of the direct and the inverse cascade of magnetic helicity at steady state can be quantified by the rate of dissipation 
in large scales $\veps_{H_b}^-$ and small scales $\veps_{H_b}^+$ as 
\begin{equation}
 \veps_{H_b}^\pm \equiv \nu^\pm \avg{\bm a \cdot \Delta^{\pm n} \bm b}.
\end{equation}
Then, the total magnetic helicity dissipation can be computed as the sum of the dissipation at large and small scales, viz., $\veps_{H_b} = \veps_{H_b}^- + \veps_{H_b}^+$. In Fig. \ref{fig:epshb}, we plot the time-series of the ratio $\veps_{H_b}^-/\veps_{H_b}$ for the three scale separations that we simulated (see Table \ref{tbl:simulations}). The values of $\veps_{H_b}^-/\veps_{H_b}$ can range from 0 to 1. For $\veps_{H_b}^-/\veps_{H_b} = 0$ there is no inverse cascade of magnetic helicity, while for $\veps_{H_b}^-/\veps_{H_b} = 1$ there is no direct cascade of magnetic helicity. 
In the initial transient regime the amount of magnetic helicity that is transferred to the large scales increases monotonically until large scale dissipation takes place and the ratio $\veps_{H_b}^-/\veps_{H_b}$ saturates to the value of 0.9 (see Fig. \ref{fig:epshb}). This clearly shows that there is a bidirectional cascade of $H_b$ with 90\% of the magnetic helicity cascading to the large scales at steady state while 10\% of the magnetic helicity cascades towards the small scales since $\veps_{H_b}^+/\veps_{H_b} \simeq 0.1$. 
The bidirectional cascade of Hb had been observed qualitatively
in Refs.~\cite{Alexakis06,Mueller12} and it was quantified in 
Ref.~\cite{ld16}.
As scale separation increases the amount of magnetic helicity that cascades to the large and small scales remains fixed in contrast to the cascade of the total energy, where $\veps^-/\veps \propto (k_fL)^{-1}$ as it was shown in \cite{ld16}, with the total energy dissipation rate being 
$\veps = \veps^- + \veps^+$ and $\veps^\pm \equiv \eta^\pm \avg{|\nabla^{\pm n} \bm b|^2} + \nu^\pm \avg{|\nabla^{\pm n}\bm u|^2}$.
Thus, for $k_fL \gg 1$ we expect the ratio $\veps^-/\veps \rightarrow 0$, implying that the total energy will cascade only %forward 
toward the small scales while the direct cascade of magnetic helicity will not vanish because $\veps_{H_b}^+/\veps_{H_b}$ is expected to 
remain finite as our numerical simulations suggest.

However, for fixed $k_fL$ it is plausible that the direct cascade of magnetic helicity vanishes in the high-magnetic-Reynolds-number limit.
This can be inferred by considering the following upper bound derived in Ref. \cite{da15} for the magnitude of the total dissipation rate of magnetic helicity at small scales for the MHD equations that involve dissipation terms with Laplacian operators (i.e., the exponent $n=1$)
\begin{equation}
\label{eq:Hb_bound}
|\veps_{H_b}| \equiv \eta |\langle \vec{a} \cdot \Delta \vec{b} \rangle|
= \eta |\avg{\vec{b} \cdot \vec{j}}|
\leq \eta \avg{|\vec{b}|^2}^{1/2} \avg{|\vec{j}|^2}^{1/2}
\leq \eta^{1/2} \avg{|\vec{b}|^2}^{1/2} \veps^{1/2} \ ,
\end{equation}
where $\veps = \eta \avg{|\vec{j}|^2} + \nu \avg{|\vec{\omega}|^2}$ is the
total energy dissipation rate in this case, $\eta$ is the magnetic resistivity,
and $\nu$ is the kinematic viscosity.  In high-Reynolds-number MHD turbulence,
$\veps$ becomes finite and independent of $\eta$ and $\nu$
\cite{Mininni09,Dallas14b,Linkmann17a}. Thus, in the limit of $\eta \to 0$
Eq.~\eqref{eq:Hb_bound} suggests that $|\veps_{H_b}| \to 0$, unless
$\avg{|\vec{b}|^2} \propto 1/\eta$, implying infinite magnetic energy.

\begin{figure}[!ht]
  \includegraphics[width=0.5\textwidth]{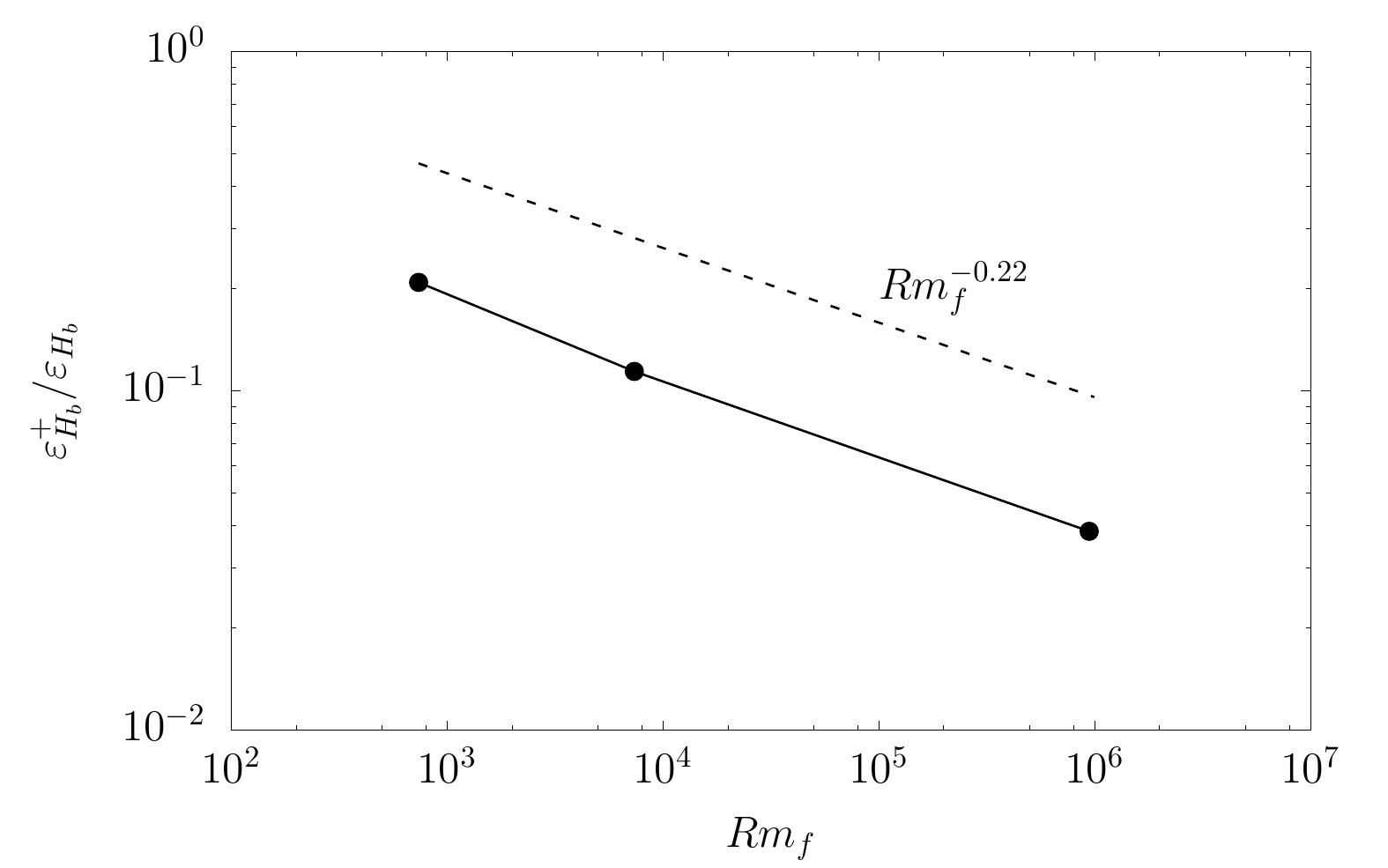}
 \caption{Dependence of $\veps_{H_b}^+/\veps_{H_b}$ on the magnetic Reynolds number $Rm_f$ for the flows with scale separation $k_fL =10$.}
 \label{fig:epsplusHb}
\end{figure}

A similar inequality for the dissipation rate of magnetic helicity at small scales for Eqs. \eqref{eq:mhd}, which involve higher-order dissipation terms, can be derived as 
\begin{equation}
\label{eq:Hb_bound_hypv}
|\veps_{H_b}^+| \equiv \eta^+ |\langle \vec{a} \cdot \Delta^{n} \vec{b} \rangle| 
               = \eta^+ |\langle \vec{b} \cdot \Delta^{n-1} \vec{j} \rangle| 
                \leqslant \eta^+ \avg{|\vec{b}|^2}^{1/2} \langle |\Delta^{n-1} \vec{j}|^2 \rangle^{1/2}.
\end{equation}
Note that for $n=1$ we recover Eq.~\eqref{eq:Hb_bound}, however, for $n>1$ the
term $\avg{|\Delta^{n-1} \vec{j}|^2}$ cannot be directly related to the
small-scale dissipation rate of magnetic energy $\veps_b^+ \equiv \eta^+
\avg{\bm b \cdot \Delta^n \bm b} = \eta^+ \avg{|\nabla^{n-1} \vec{j}|^2}$.
Therefore, to make any conclusions, if any, for the behavior of
$|\veps_{H_b}^+|$ in the limit of $\eta^+ \rightarrow 0$, we need to compute
the scaling of $\veps_{H_b}^+$ with $\eta^+$ as we cannot make any \textit{a
priori} statements like in Eq.~\eqref{eq:Hb_bound}.  For this purpose, we
performed three simulations with the same scale separation $k_fL = 10$ but
different $Rm_f$. These preliminary results are plotted in Fig.
\ref{fig:epsplusHb}, which shows that $\veps_{H_b}^+$ normalized by the total
dissipation of magnetic helicity $\veps_{H_b}$ has a weak power-law behavior in
terms of $Rm_f$, i.e. $\veps_{H_b}^+/\veps_{H_b} \propto Rm_f^{-0.22}$. This
scaling suggests that the direct cascade of magnetic helicity may not persist
in the limit of $Rm_f \to \infty$. However, further investigation into this
issue is necessary in order to understand this scaling law and the small-scale
dynamics of the magnetic helicity.

\pagebreak

\section{Spectral dynamics} \label{sec:spectral_dynamics}

\begin{figure}[!ht]
\begin{subfigure}{0.49\textwidth}
  \includegraphics[width=\textwidth]{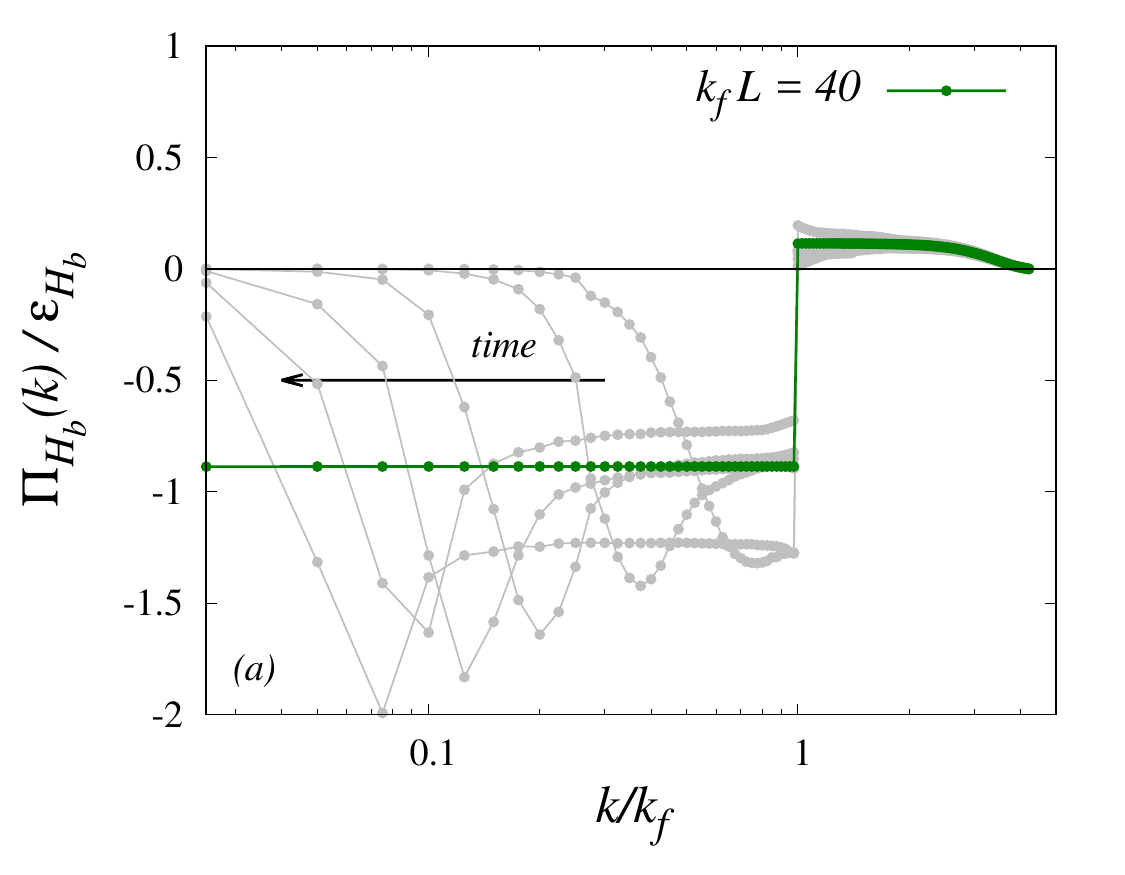}
 \caption{}
 \label{fig:fluxhb}
\end{subfigure}
\begin{subfigure}{0.49\textwidth}
  \includegraphics[width=\textwidth]{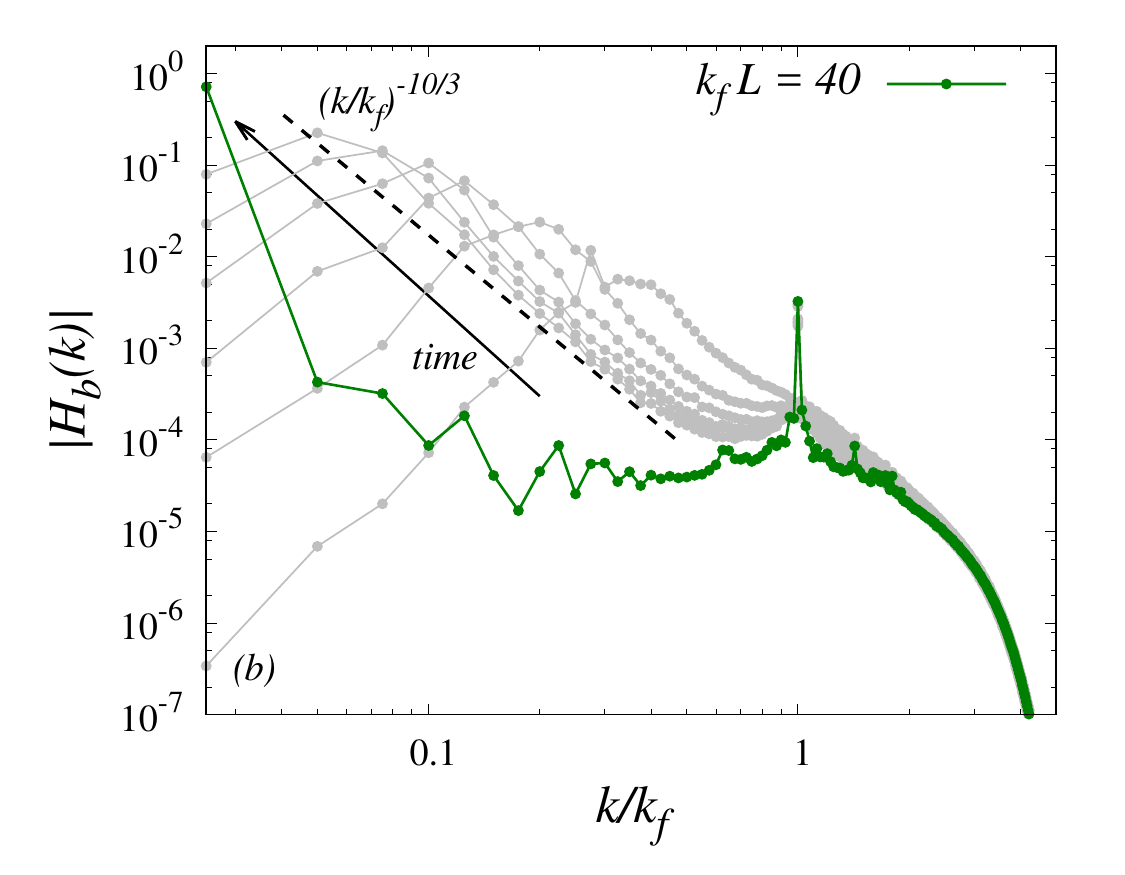}
 \caption{}
 \label{fig:spechb}
\end{subfigure}
 \caption{ 
(a) Magnetic helicity flux $\Pi_{H_b}(k)$ normalized by $\veps_{H_b}$ and 
(b) absolute value of the magnetic helicity spectrum $|H_b(k)|$ for a flow with scale separation $k_fL = 40$. 
The light gray curves denote instantaneous quantities at the transient 
regime of the flow {for $0.5\leqslant t/t_f \leqslant 3$}, 
while the green (dark gray) curve denotes the time-averaged profile in the steady-state regime.
}%\blue{\cite{ld16}}.}
 \label{fig:spectra}
\end{figure}

After %looking at 
considering
the global dynamics and the time evolution of our flows, we now elaborate on the spectral dynamics of magnetic helicity in order to have a clear picture of the dynamics across scales. 
This is best demonstrated by looking at the spectrum of magnetic helicity $H_b(k)$ and its normalized flux $\Pi_{H_b}(k)/\veps_{H_b}$ for the flow with scale separation $k_fL = 40$ (see Fig. \ref{fig:spectra}). The magnetic helicity flux is defined as
%q
\begin{equation}
\Pi_{H_b}(k)= \sum_{k'=1}^k 
               \sum_{|\vk|=k'} \hat \vb_{\vk}^* \cdot \widehat{(\vu \times \vb)}_{\vk} \ ,
\end{equation}
and the magnetic helicity spectrum as $H_b(k)= \sum_{|\vk|=k} \hat \va_{\vk}^*
\cdot \hat \vb_{\vk} $, where the caret denotes the Fourier modes of the
corresponding real vector fields and the asterisk the complex conjugate.
The gray curves denote instantaneous profiles at the transient regime of the flow while the green (dark gray) curves denote the time-averaged profiles at the steady state regime.
Negative values of the magnetic helicity flux imply upscale transfer %inverse cascades 
while positive values imply %forward cascades 
downscale transfer.
Thus, Fig. \ref{fig:fluxhb} shows clearly the bidirectional cascade behavior of magnetic helicity with most of $H_b$ cascading towards large scales in agreement with Fig. \ref{fig:epshb}. The gray curves indicate how the flux builds up at the transient state and of course these curves do not satisfy the balance $\Pi_{H_b}(k) = \veps_{H_b}$, which is only expected to be valid on average over a statistically stationary regime.

\begin{figure}[!ht]
 \begin{subfigure}{0.49\textwidth}
 \includegraphics[width=\textwidth]{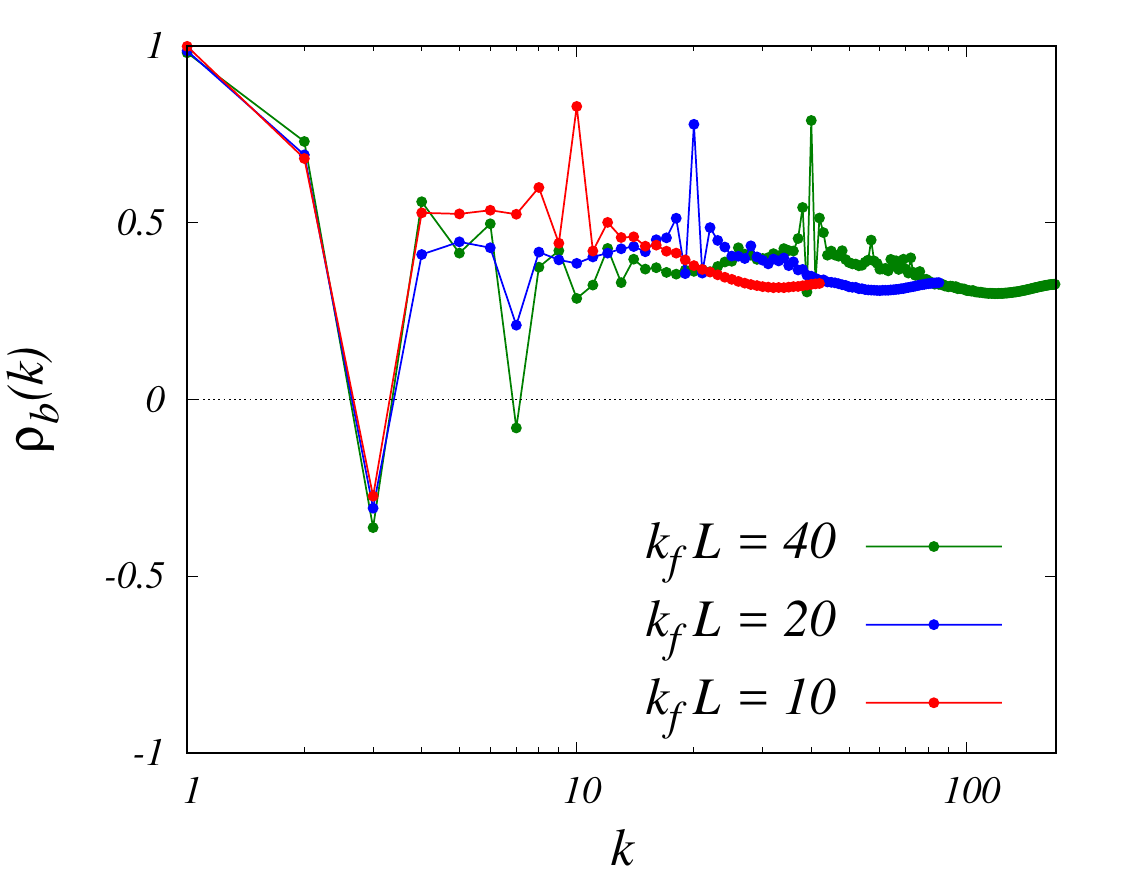}
 \caption{}
\label{fig:rhobk}
\end{subfigure}
\begin{subfigure}{0.49\textwidth}
 \includegraphics[width=\textwidth]{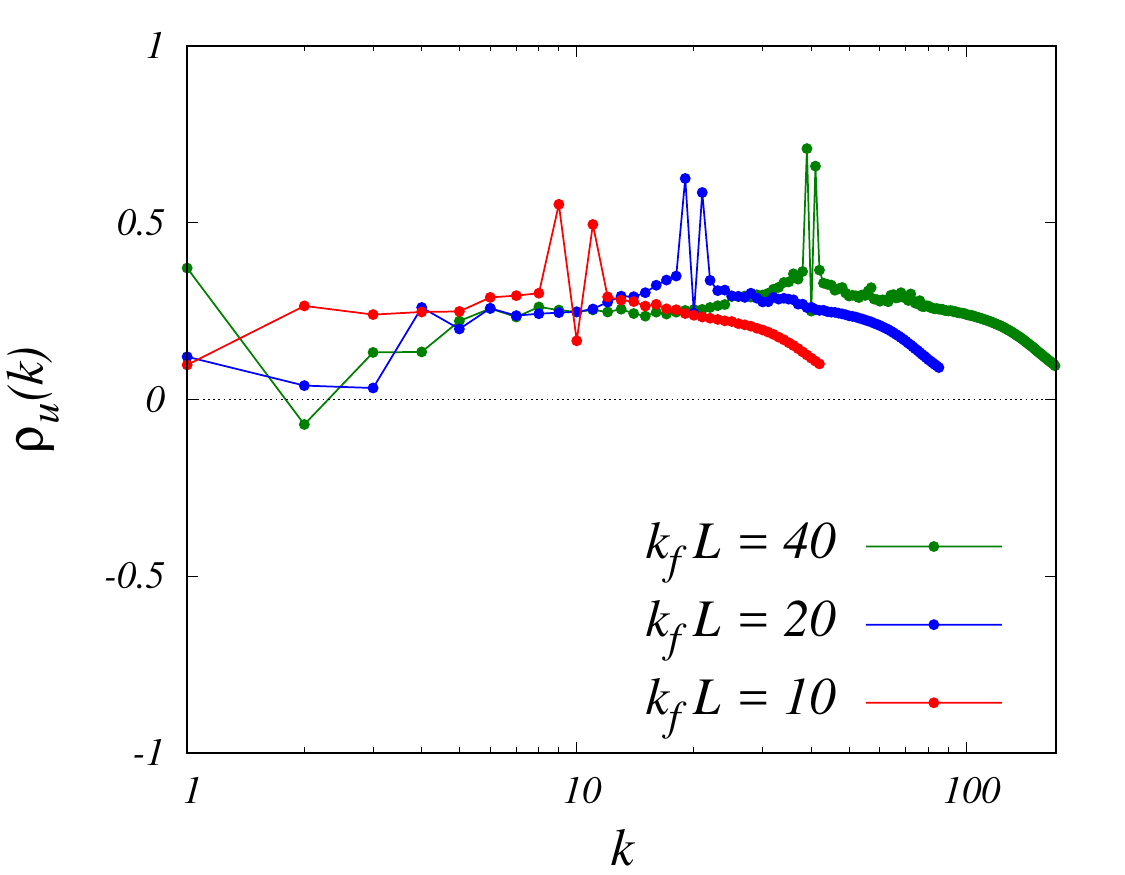}
 \caption{}
\label{fig:rhouk}
\end{subfigure}
 \caption{
(a) Time-averaged relative magnetic helicity spectra $\rho_b(k)$ and
(b) time-averaged relative kinetic helicity spectra $\rho_u(k)$ for flows with scale separations $k_fL = 10, 20$, and 40.
}
 \label{fig:rhok}
\end{figure}

At the transient stage of the simulation the inverse and the direct cascade of
magnetic helicity is dominated by local interactions, as it has already been
reported \cite{Alexakis06}. Before the flow saturates to a steady-state
solution a spectrum with a negative power-law slope can be observed at low
wave numbers. The scaling of this spectrum was proposed to be $H_b(k) \propto
k^{-10/3}$ \cite{Mueller12,Mininni09}, which agrees with our data. However, if
one integrates further in time the spectrum develops even more as the flow
saturates to a statistically stationary regime. In this regime, most of the
magnetic helicity is concentrated at the largest scales [see also the spectrum
of normalized helicity $\rho_b(k)$ in Fig. \ref{fig:rhok}(a)] and the inverse
cascade of $H_b$ becomes nonlocal in wave-number space, i.e., $H_b$ is
transferred directly from the forced scale to the largest scales of the flow,
while the forward cascade remains local \cite{ld16,Alexakis06,Brandenburg01}. 
{This transition to nonlocal interactions at steady state occurs when
condensation takes place at the largest scale of the system. In this case, the
large-scale vortex interacts directly with the small-scale vortices of the flow
(i.e., nonlocally in wave-number space) \cite{ld16}. This behavior is also
observed in confined two-dimensional hydrodynamic turbulent flows
\cite{shuklaetal16,xiaetal08}}.
Finally, the scaling of the time-averaged spectrum of magnetic helicity at
steady state is $H_b(k) \propto k^0$ for wave numbers $k<k_f$, suggesting that
the large scales behave as if they were in absolute equilibrium \cite{ld16}.
Now, the scaling for $k>k_f$ seems to be $H_b(k) \propto k^{-5/3}$. However,
this is not conclusive from the plot due to the limited spectral range and the
high-order dissipative terms that act at the largest wave numbers.

\begin{figure}[!ht]
  \includegraphics[width=0.5\textwidth]{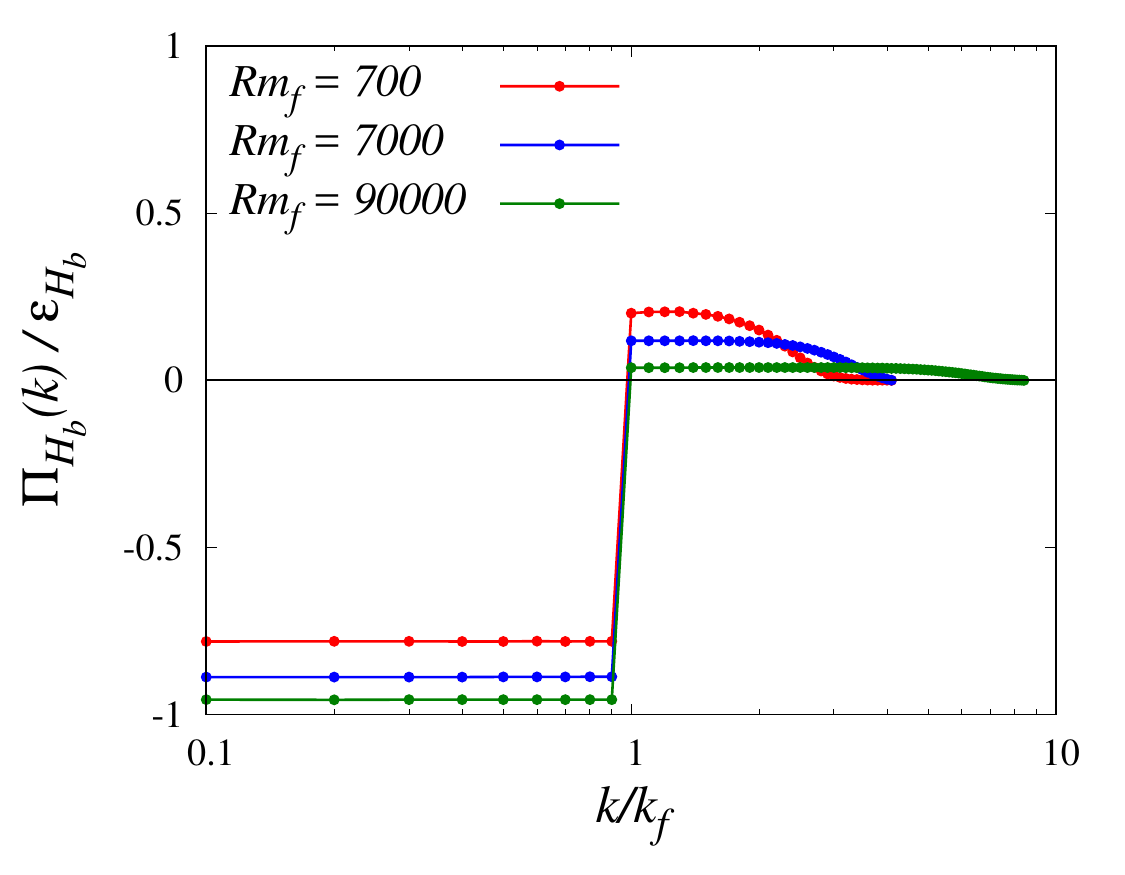}
 \caption{
{
Spectral flux of magnetic helicity normalised by its dissipation rate $\Pi_{H_b}(k)/\veps_{H_b}$ at different values of $Rm_f$ for the flows with scale separation $k_fL =10$. %The height of the plateau at 
The amplitude of the cascade at $k/k_f > 1$ is quantified as a function of $Rm_f$ in Fig.~\ref{fig:epsplusHb}.
}}
 \label{fig:epsplusHb_flux}
\end{figure}

{The behavior of the magnetic helicity flux $\Pi_{H_b}(k)$ at steady-state
with increasing $Rm_f$ is presented in Fig.~\ref{fig:epsplusHb_flux}. 
As can be seen from the figure, 
the small-scale range where $\Pi_{H_b}(k) \simeq \veps_{H_b}^+$ 
extends over a larger range of wave numbers successively
with increasing $Rm_f$ as expected for a cascade process. 
However, the height of the plateau, i.e., the value of 
$\veps_{H_b}^+$, decreases with increasing $Rm_f$, as quantified in Fig.~\ref{fig:epsplusHb}. 
}

The purpose of our simulations using a positively helical electromagnetic force
was to impose a mean magnetic helicity in our flows with the intention to have
a dominant sign of magnetic helicity across the scales. However, at the
steady-state regime of our simulations there are low-wave-number modes that
develop a negative sign of magnetic helicity even though the mean value of
magnetic helicity is positive in the flow. This is clearly depicted in Fig.
\ref{fig:rhobk}, where we plot the time-averaged spectra of the normalized
magnetic helicity $\rho_b(k)$ for flows with $k_fL = 10, 20, 40$.
Note that all the flows generate negative $\rho_b(k)$ at wave number $k = 3$, which indicates that there is a consistent mechanism generating opposite signs of helicity across the scales even in these flows that are dominated by positive net magnetic helicity. Moreover, we observe that the number of modes at $k<k_f$ with negative magnetic helicity increases as the scale separation $k_fL$ increases.

{Despite the nonhelical mechanical forcing, a whole spectrum of positive kinetic
helicity is observed across the scales.} This is shown in Fig. \ref{fig:rhouk},
where we plot the spectra of the relative kinetic helicity $\rho_u(k) \equiv
H_u(k) / (\avg{|{\hat\vu}_{\vk}|^2}\avg{|{\hat{\bm\omega}}_{\vk}|^2})^{1/2}$
for the flows with $k_fL = 10, 20,$ and $40$.  {The values of $\rho_u(k)$ for
wave numbers $k<k_f$ decrease as the scale separation $k_fL$ increases, while
they are particularly high} close to the forcing scale where $\rho_b(k)$ is
also dominant. This correlation seems to be related to the injection of
positive magnetic helicity in the flow. In order to understand further the
observations from our numerical simulation on the generation of the $\rho_u(k)$
spectrum, the direct cascade of magnetic helicity, and the mechanism that
generates opposite signs of magnetic helicity between large and small scales,
we study analytically the triadic interactions of helical modes in the
following section.

\section{Triad interactions of helical modes} \label{sec:triadic_analysis}
\begin{figure}[!ht]
\center
\includegraphics[scale=0.12]{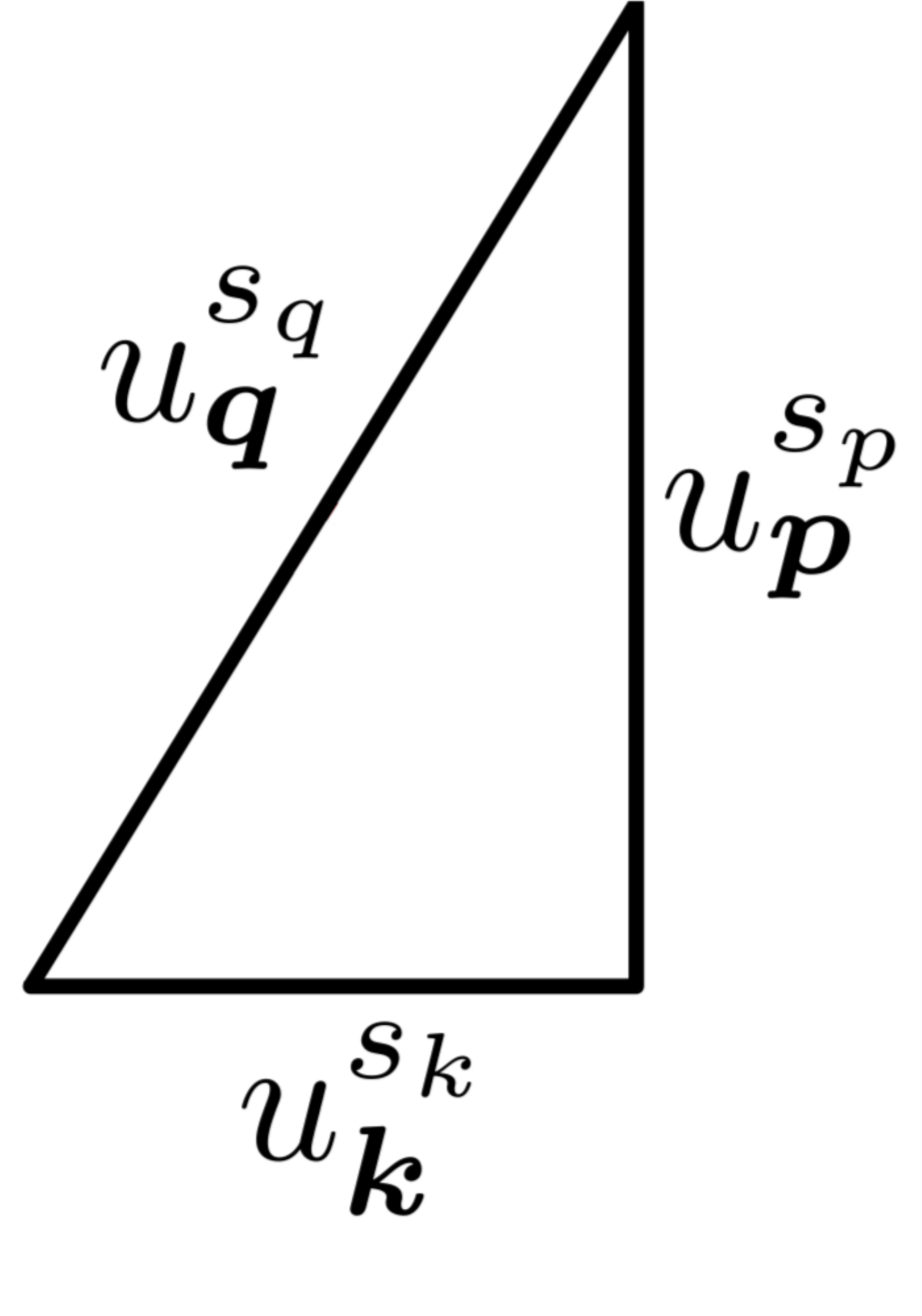}
\includegraphics[scale=0.12]{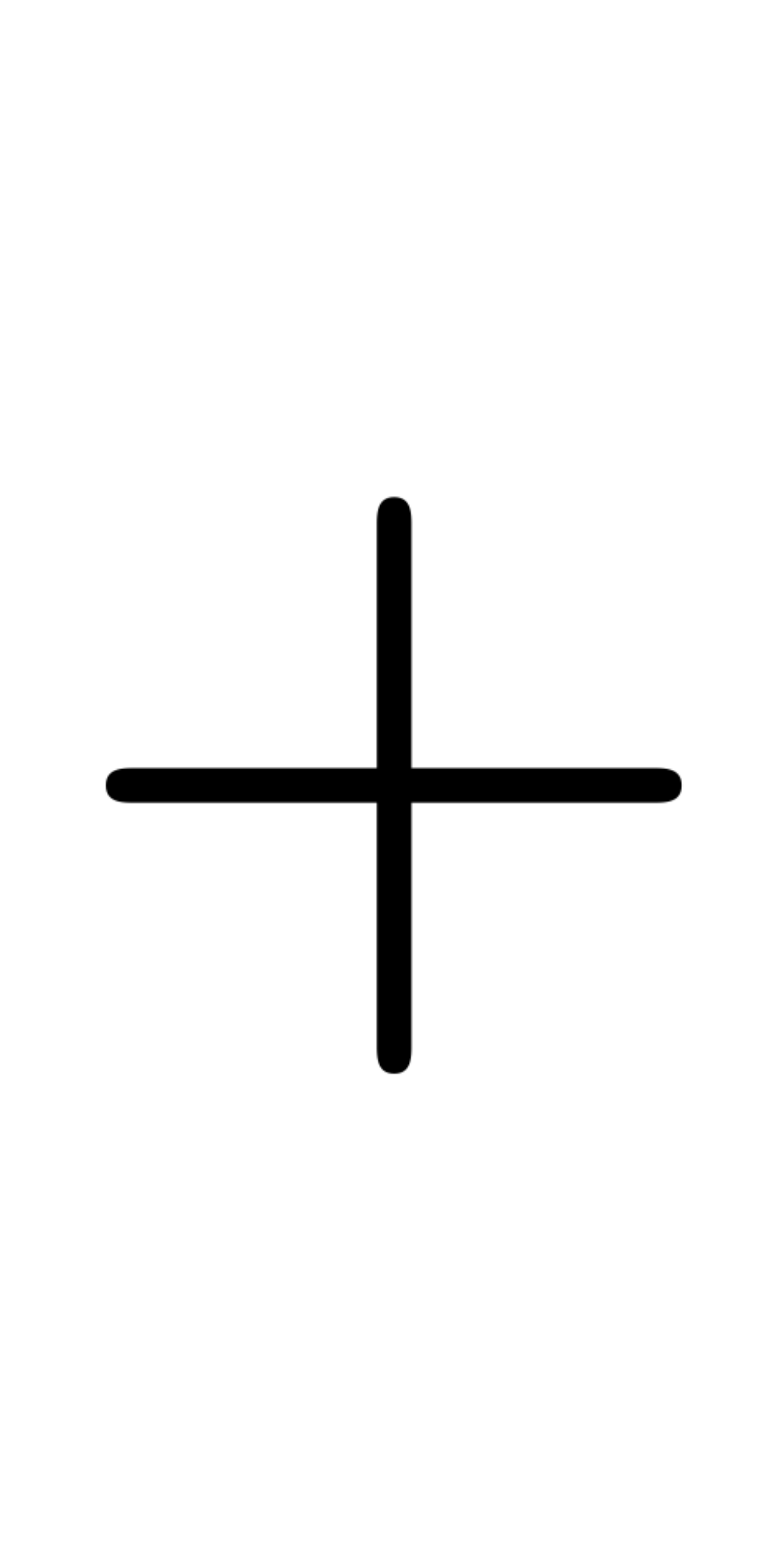}
\includegraphics[scale=0.12]{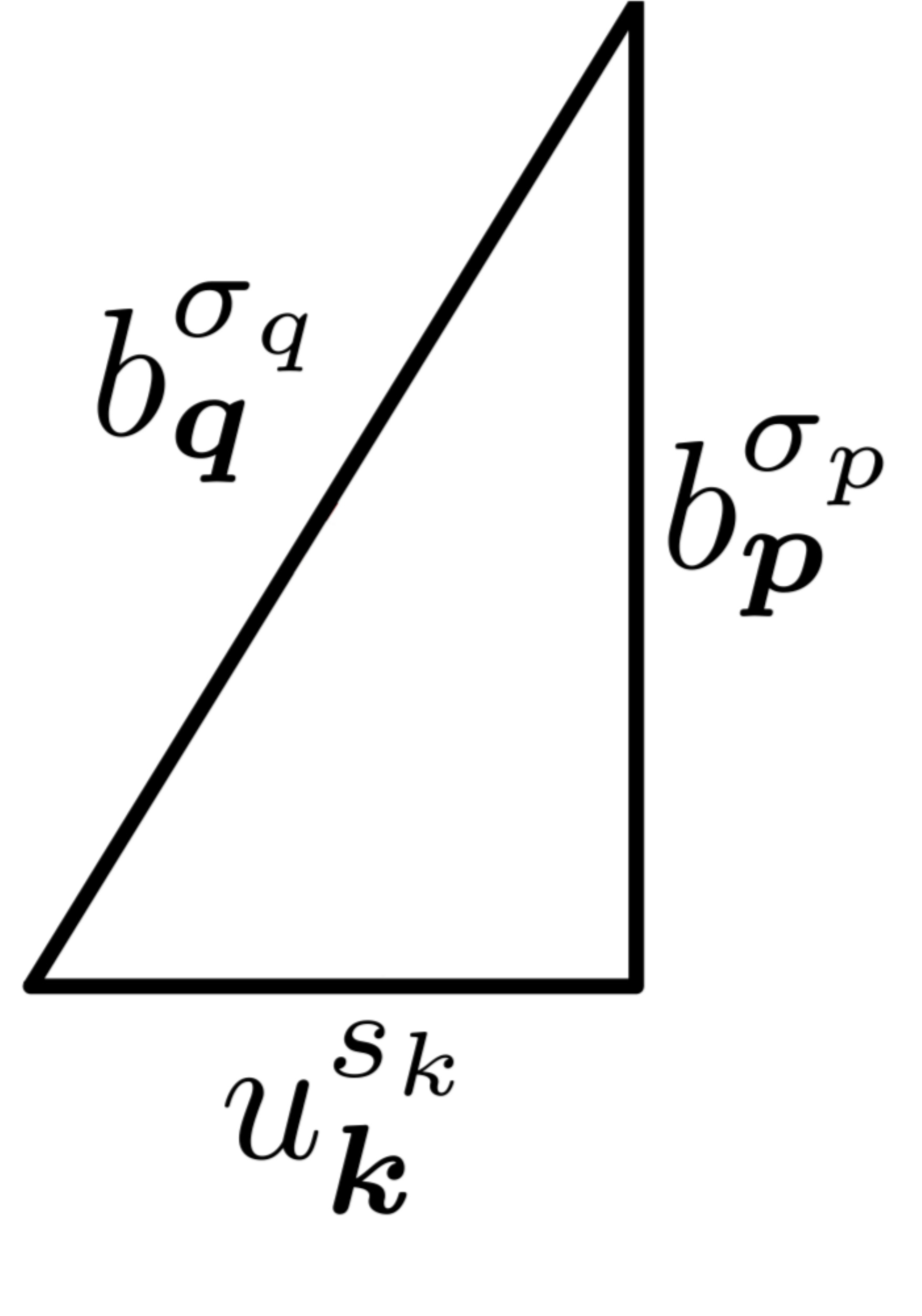}
\includegraphics[scale=0.12]{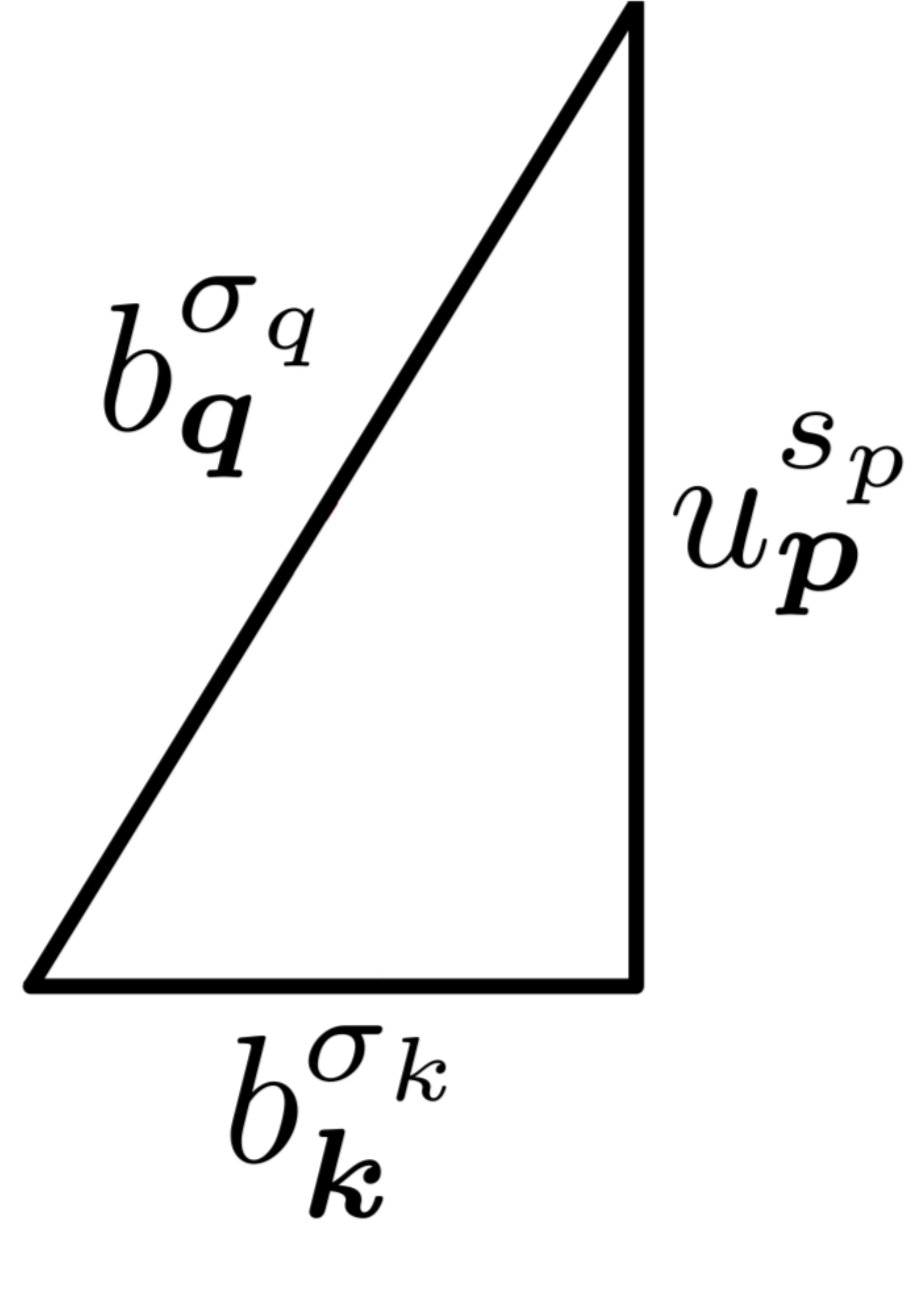}
\includegraphics[scale=0.12]{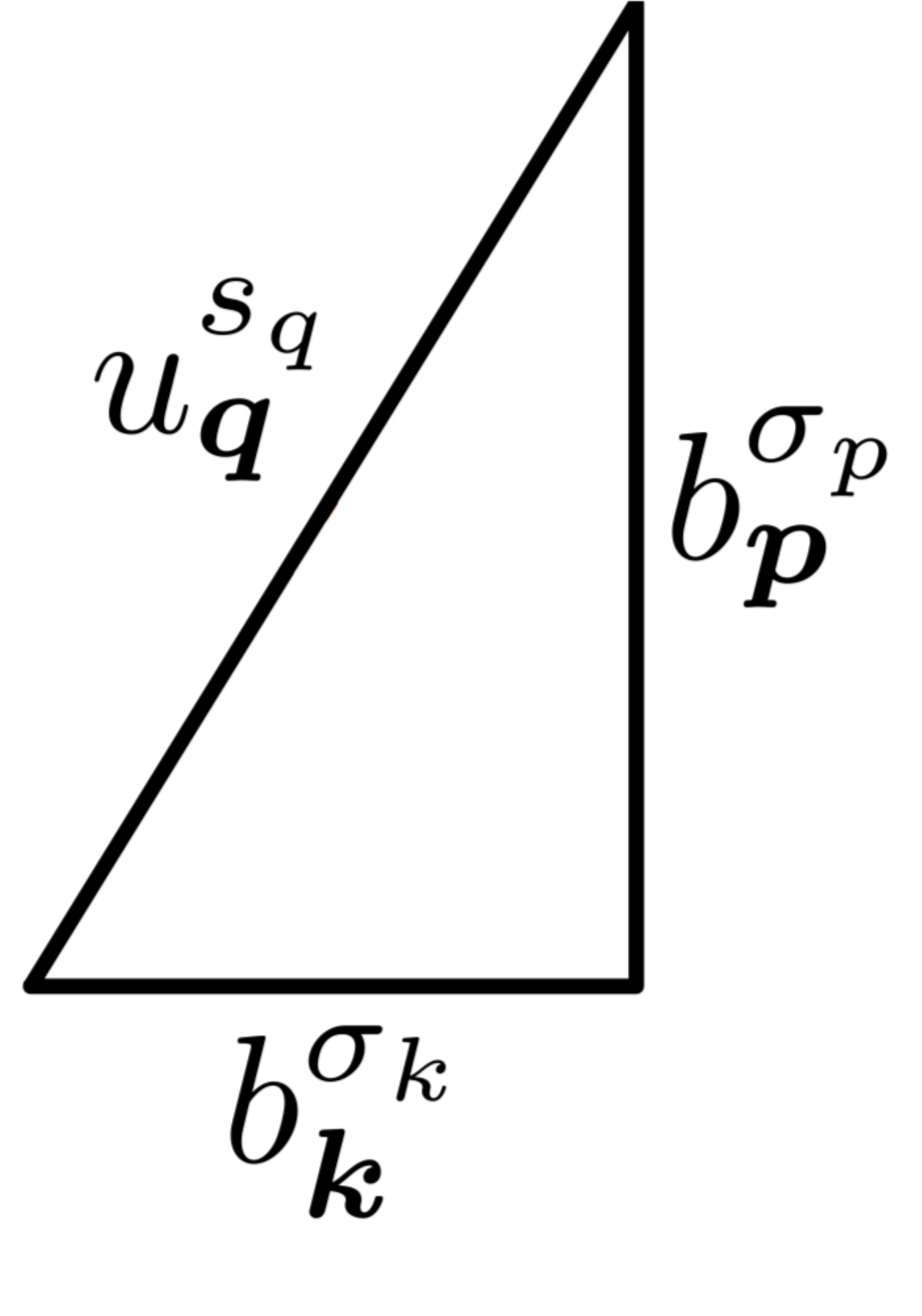}
\caption{
Graphical representation of the dynamical system given by
Eqs.~\eqref{eq:basic-triads} consisting of the coupling among velocity (left)
and magnetic-velocity (right) triadic interactions as in Ref.~\cite{Linkmann17b}. 
The magnetic-velocity
interactions correspond to several terms in Eqs.~\eqref{eq:basic-triads},
depending on whether the time evolution of the velocity or the magnetic field
modes is considered.
}
\label{fig:triads_graphical}
\end{figure}

The three-dimensional vector fields of the velocity and the magnetic field are
solenoidal 
(viz., $i \bm k \cdot {\hat{\bm u}}_{\vk} = 0$ and $ i \bm k \cdot {\hat{\bm b}}_{\vk} = 0$), 
hence ${\hat{\bm u}}_{\vk}$ and ${\hat{\bm b}}_{\vk}$ must lie in the plane perpendicular to the wave vector $\vk$. 
This plane is spanned by two eigenvectors of the curl
operator in Fourier space with nonzero eigenvalues. Since these eigenvectors
are by definition fully helical, each Fourier mode ${\hat{\bm u}}_{\vk}$
and ${\hat{\bm b}}_{\vk}$ can be further decomposed into two modes with
positive and negative helicity
\begin{align}
\label{eq:basisu}
{\hat{\bm u}}_{\vk}(t)&=\ukp(t) \hkp + \ukm(t) \hkm=\sum_{s_k} \usk(t) \hsk \ , \\
\label{eq:basisb}
{\hat{\bm b}}_{\vk}(t)&=\bkp(t) \hkp + \bkm(t) \hkm=\sum_{\sigma_k} \bski(t) \hski \ ,
\end{align}
where the basis vectors $\vec{h}^{\pm}_{\vk}$
are the 
orthonormal eigenvectors of the curl operator in Fourier space satisfying
$i\vk \times \hsk = s_k |\bm k| \hsk$ with $s_k = \pm$ and $\sigma_k = \pm$. 
This is the so-called {helical decomposition} \cite{Craya58,Lesieur72,Herring74}.
To study the triad interaction of helical modes, we use the formalism that was
developed by Waleffe \cite{Waleffe92} for the Navier-Stokes equations and extended by 
Linkmann {\em et al.}~\cite{Linkmannetal16,Linkmann17b} 
to the MHD equations.
This is essentially a linear stability analysis of a 
dynamical system obtained from the MHD equations in the limits of $\nu \to 0$ and $\eta \to 0$.
In the MHD equations that are shown here in Fourier space \\
\begin{align}
\label{eq:Fmomentum}
(\dt + \nu k^2) \fvec{u}_{\vk} = & -\mathcal{F}\left[\nabla \left( p + \frac{|\vu|^2}{2}\right) \right] 
  +\sum_{\vec{k}+\vec{p}+\vec{q}=0} \left( (-i{\vp}\times \fvec{u}_{\vp})^*
       \times \fvec{u}_{\vq}^*+(i\vec{p}\times \fvec{b}_{\vp})^*\times \fvec{b}_{\vq}^*\right)  \ , \\
\label{eq:Finduction}
(\dt + \eta k^2) \fvec{b}_{\vk} = & \  i\vec{k} \times \sum_{\vec{k}+\vec{p}+\vec{q}=0} \fvec{u}_{\vp}^* \times \fvec{b}_{\vq}^* \ , 
\end{align} 
with $\mathcal{F}$ denoting the Fourier transform as a linear operator,
the convolutions that describe the respective Fourier transforms of
the inertial term $\vu \times \vw$, the Lorentz force $\vec{j} \times \vb$, and
the curl of the electromotive force $\nabla \times (\vu \times \vb)$ 
are
reduced to single triads of wave vectors $\vk$, $\vp,$ and $\vq$
satisfying $\vk + \vp +\vq =0$. For convenience and without loss of generality we impose the
ordering $|\vk| \leq |\vp| \leq |\vq|$. 
In order to study the dynamics in the inertial range of scales, 
the dissipation terms are neglected 
in the limits of $\nu \to 0$ and $\eta \to 0$.
Following Lessinnes \etal~\cite{Lessinnes09}, 
by substituting Eqs.~\eqref{eq:basisu} and \eqref{eq:basisb} into 
Eqs.~\eqref{eq:Fmomentum} and \eqref{eq:Finduction} after restricting the convolutions to single triads
and subsequently taking the inner product with the helical basis vectors,
one can then derive the following system of ordinary differential 
equations, 
which conserves the ideal invariants of the MHD equations \cite{Lessinnes09}:
\begin{align}
\label{eq:basic-triads}
\dt {\usk}^* &=  g_{s_k s_p s_q} (s_pp-s_qq)\:  \usp \usq  - g_{s_k\sigma_p \sigma_q} (\sigma_pp-\sigma_qq)\:  \bspi \bsqi \ , \nonumber \\
\dt {\usp}^* &=  g_{s_k s_ps_q} (s_qq-s_kk)\:  \usq \usk - g_{\sigma_ks_p\sigma_q}(\sigma_qq-\sigma_kk)\:  \bsqi \bski \ , \nonumber \\
\dt {\usq}^* &=  g_{s_k s_p s_q} (s_kk-s_pp)\:  \usk \usp - g_{\sigma_k \sigma_ps_q} (\sigma_kk-\sigma_pp)\:  \bski \bspi \ , \nonumber \\
\dt {\bski}^* &=  \sigma_k k\: \left(g_{\sigma_k\sigma_p s_q} \bspi \usq - g_{\sigma_ks_p \sigma_q} \usp \bsqi \right) \ ,\nonumber \\
\dt {\bspi}^* &=  \sigma_p p\: \left(g_{s_k \sigma_p \sigma_q} \bsqi \usk -  g_{\sigma_k\sigma_p s_q} \usq \bski \right) \ , \nonumber  \\
\dt {\bsqi}^* &=  \sigma_q q\: \left(g_{\sigma_k s_p\sigma_q}\bski \usp -  g_{s_k \sigma_p\sigma_q}\usk \bspi \right) \ ,
\end{align}

where the geometric factors $g$ describe the coupling between the helical basis
vectors, i.e.  $g_{s_k s_p s_q} = \hsk \cdot ({\hsp} \times {\hsq})$
\cite{Waleffe92}.  Further details on the full derivation of
Eqs.~\eqref{eq:basic-triads} can be found in
Refs.~\cite{Lessinnes09,Linkmannetal16,Linkmann17b}.  A graphical
representation of the system of equations \eqref{eq:basic-triads} is shown in
Fig.~\ref{fig:triads_graphical}, where each convolution term corresponds to a
triangle that represents a triad of wave vectors.  Note that two triads are
required to describe the interactions that correspond to the term $\nabla
\times (\vu \times \vb)$ due to a necessary symmetrization of the convolution
in Fourier space.  However, this symmetrization does not allow one to disentangle
the triadic interactions of the advection term $\vec{u} \cdot \nabla \vec{b}$
and the stretching term $\bm b \cdot \nabla \bm u$ of the magnetic field
\cite{Linkmannetal16}.  Different interactions of helical modes can now be
studied via Eqs.~\eqref{eq:basic-triads} by choosing different combinations of
$s_k, s_p, s_q$ and $\sigma_k, \sigma_p, \sigma_q$.
%
%%%%%%%%%%%%%% GENERAL TRIADIC  SYSTEM %%%%%%%%%
%%%%%%%%%%%%%%

A linear stability analysis of steady solutions of Eqs.~\eqref{eq:basic-triads} has been carried out by
Linkmann {\em et al.}~\cite{Linkmannetal16},
where a linear instability corresponds to a transfer of energy
from the unstable helical mode (denoted in capital letters)
of a single triad interaction to the perturbations (denoted in lower case letters), 
i.e., to the other two helical modes of the triad. 
Under the assumption that the
statistical behavior of the flow is controlled by the stability characteristics
of these isolated triads (referred to as the instability assumption)
\cite{Kraichnan67,Waleffe92}, 
it is possible to draw conclusions concerning the
energy and helicity transfer 
in the MHD equations from 
the stability properties of Eqs.~\eqref{eq:basic-triads}. 
Different physical processes such as the inverse cascade of magnetic helicity or kinematic
dynamo action can then be studied on the level of %single 
triad interactions by setting up specific 
perturbation problems \cite{Linkmannetal16}. 
Depending on the characteristic wave numbers and the 
sign of helicities of the interacting modes, 
the dominant interscale energy transfers can then be identified through a
comparison of the growth rates of the perturbations \cite{Linkmann17b}. 

\begin{figure}[!ht]
\center
\begin{subfigure}{0.12\textwidth}
  \includegraphics[width=\textwidth]{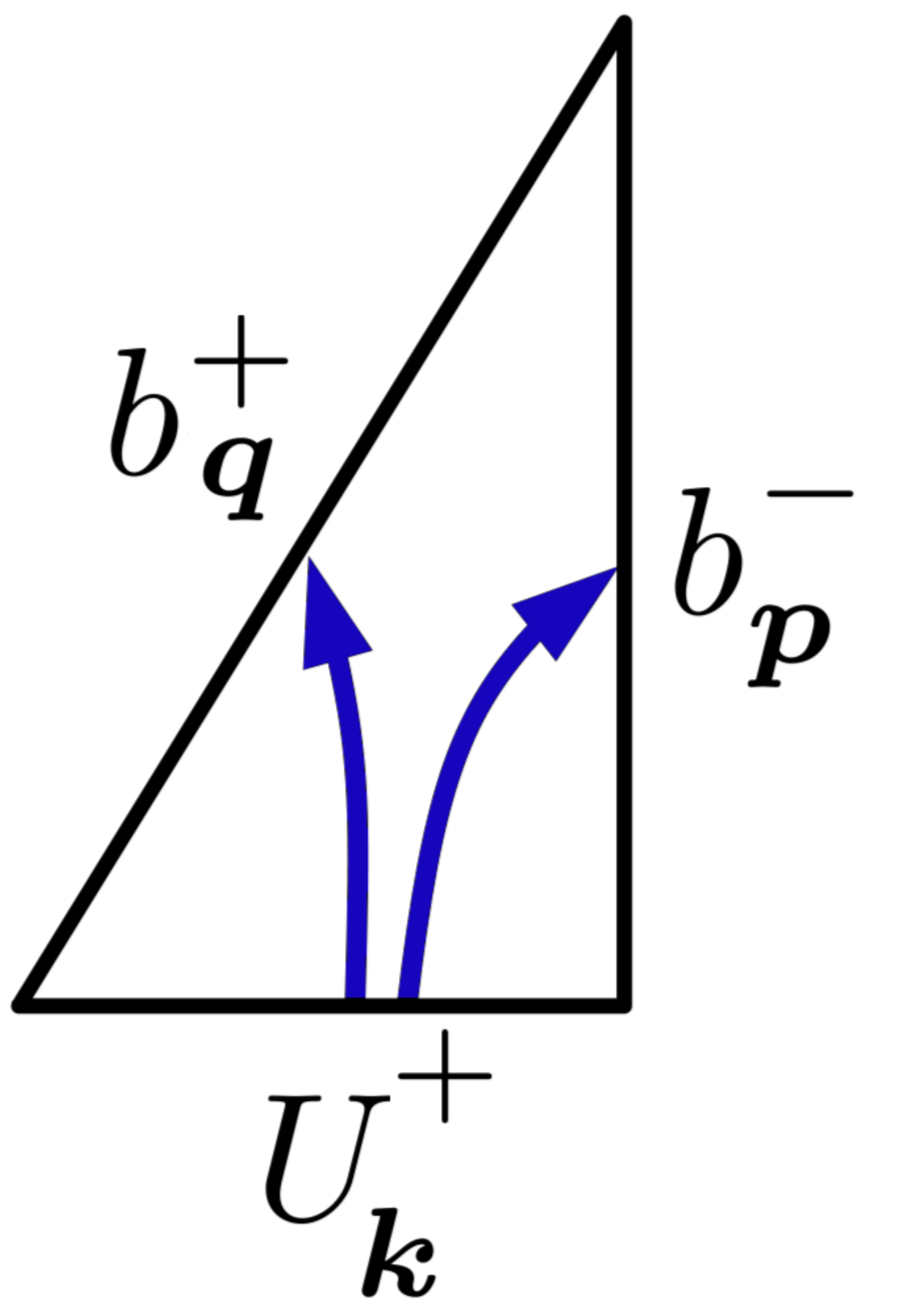}
  \caption{}
  \label{fig:5a}
\end{subfigure}
\begin{subfigure}{0.12\textwidth}
  \includegraphics[width=\textwidth]{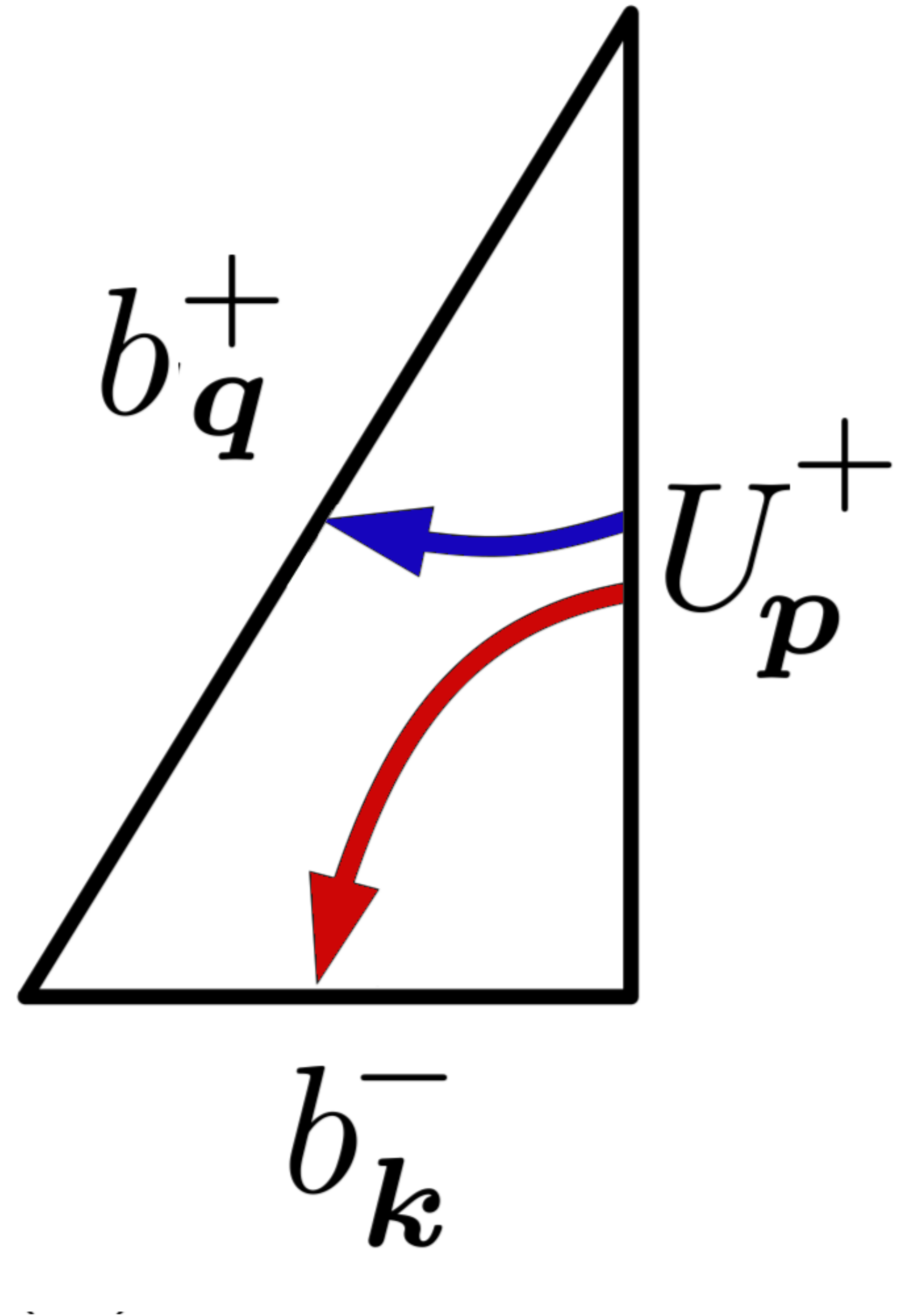}
  \caption{}
  \label{fig:5c}
\end{subfigure} 
\begin{subfigure}{0.12\textwidth}
  \includegraphics[width=\textwidth]{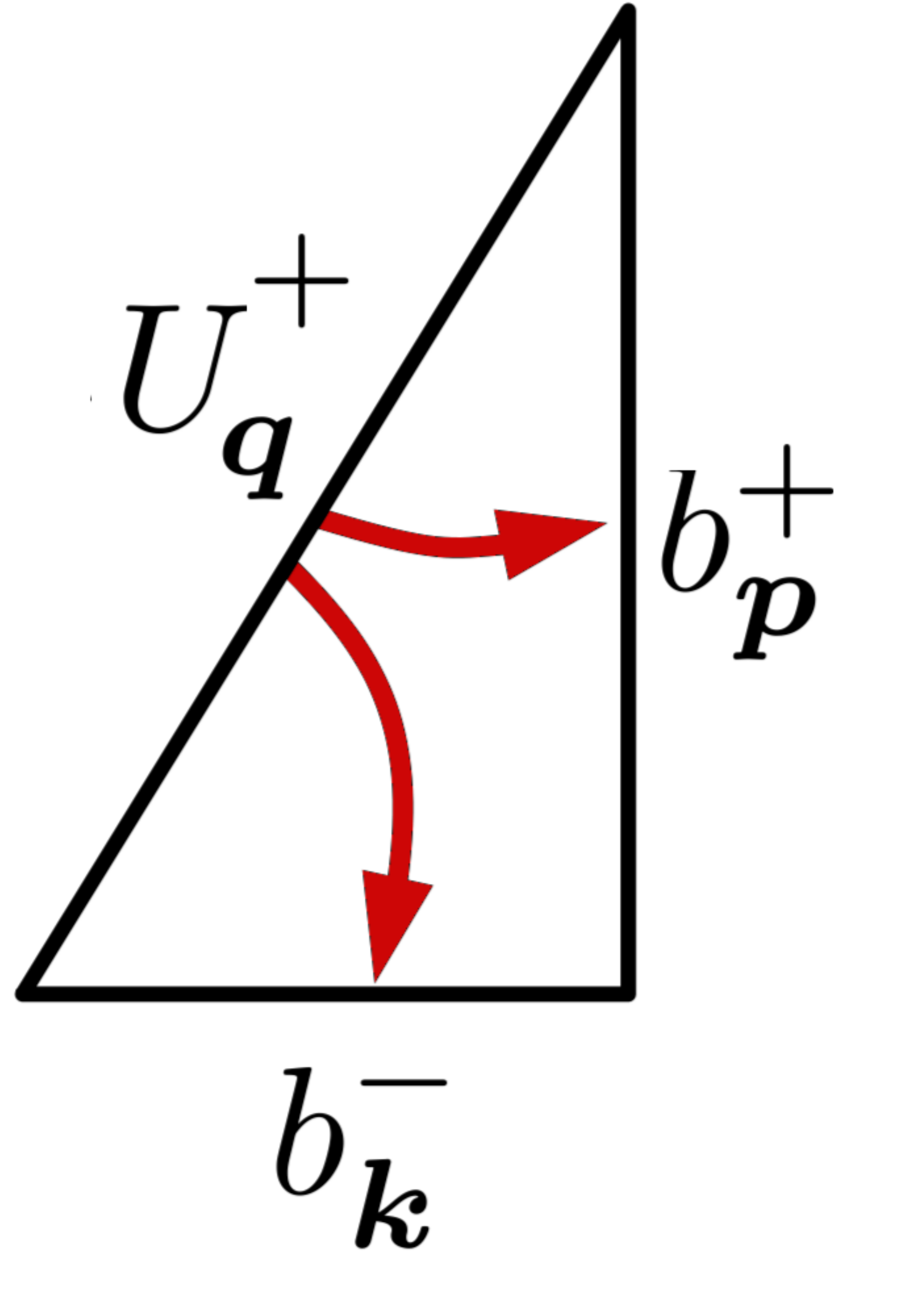}
  \caption{}
  \label{fig:5e}
\end{subfigure} 
\\
\begin{subfigure}{0.12\textwidth}
  \includegraphics[width=\textwidth]{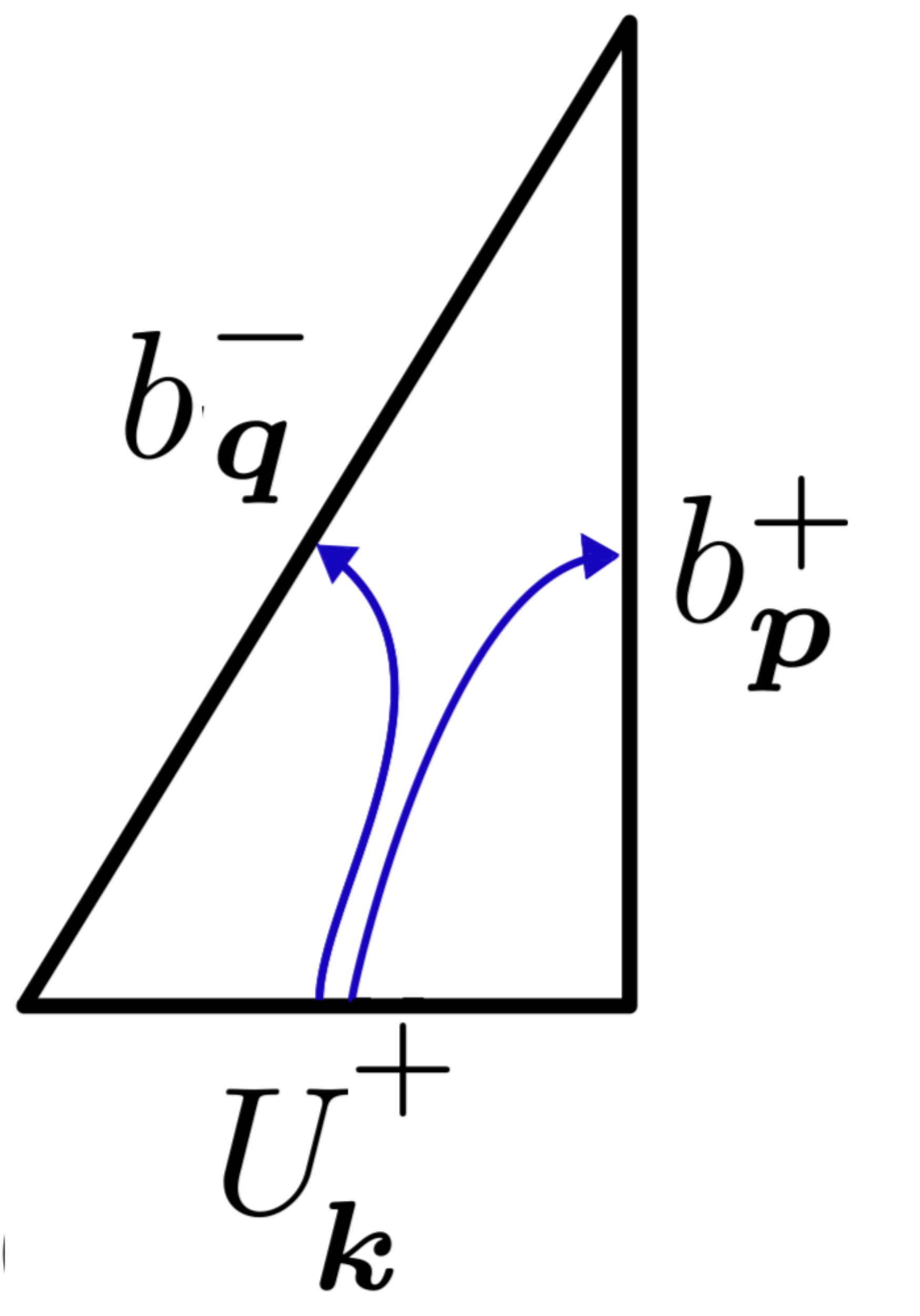}
  \caption{}
  \label{fig:5b}
\end{subfigure}
\begin{subfigure}{0.12\textwidth}
  \includegraphics[width=\textwidth]{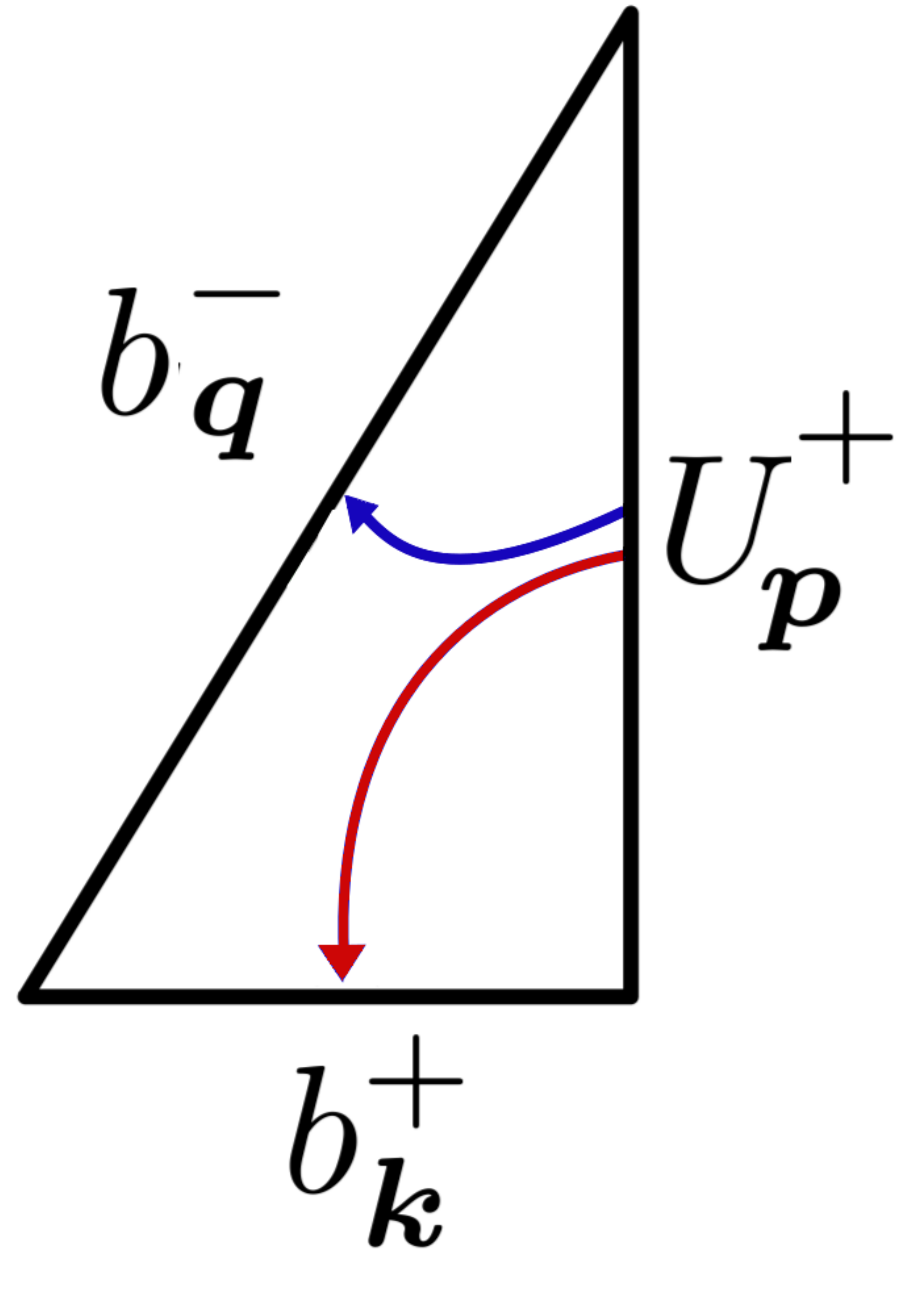}
  \caption{}
  \label{fig:5d}
\end{subfigure}
\begin{subfigure}{0.12\textwidth}
  \includegraphics[width=\textwidth]{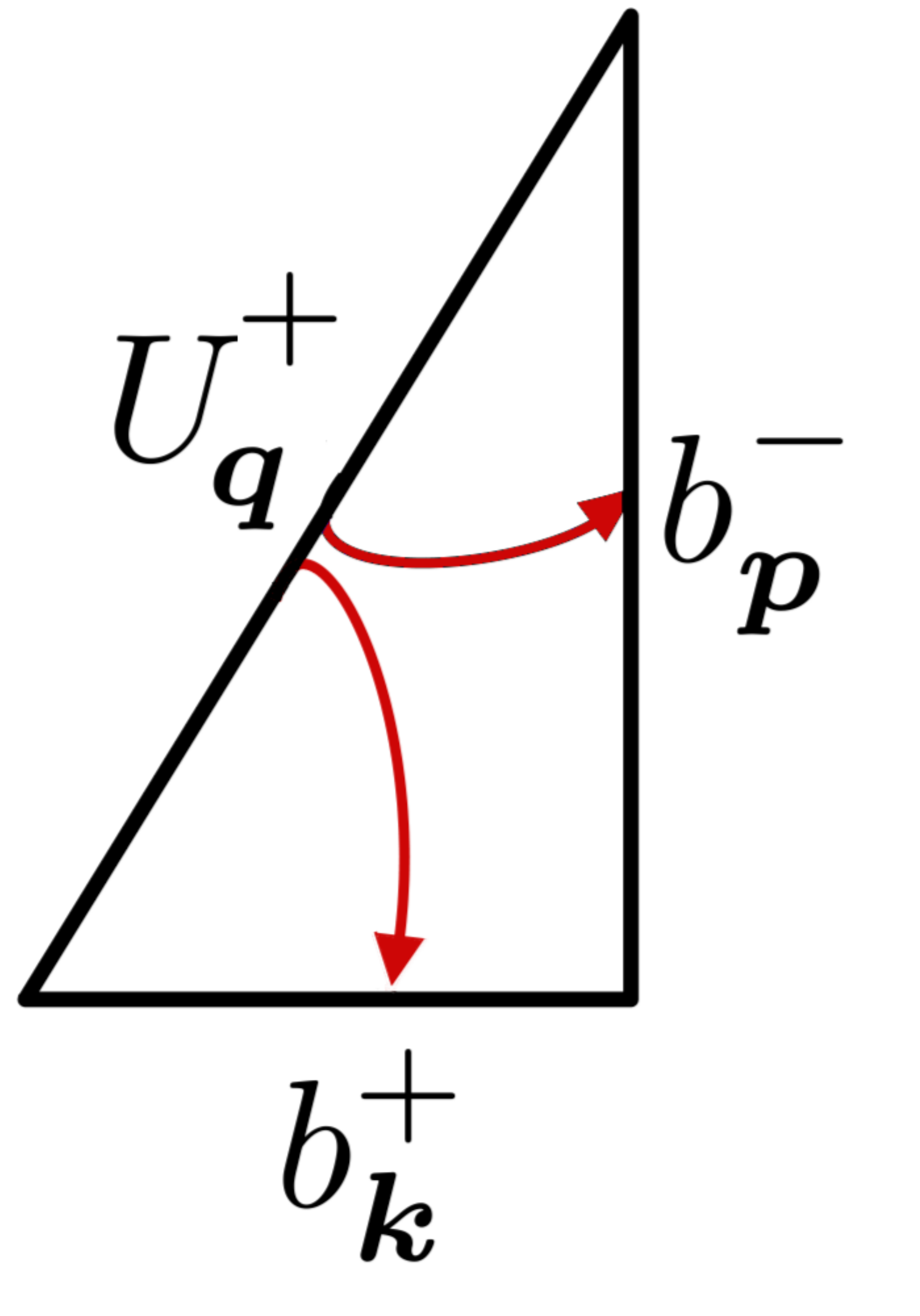}
  \caption{}
  \label{fig:5f}
\end{subfigure}
\caption{Triadic interactions of helical modes leading to energy transfers
when considering a steady solution for the velocity field 
subject to magnetic perturbations. Blue (dark gray) arrows indicate 
small-scale dynamos and 
red (light gray) arrows indicate large-scale dynamos.
The thickness of the arrows indicates the magnitude of the transfer.}
\label{fig:triadic_dynamos}
\end{figure}

If we consider a steady solution for the velocity field subject to magnetic
perturbations, it is possible to identify all the triadic
interactions of 
helical modes that produce small as well as
large scale growth of the magnetic field (see Ref.~\cite{Linkmannetal16} for details).
With this result Linkmann {\em et al.}~\cite{Linkmann17b} were able to interpret
how small- and large-scale dynamos operate at the level of triadic interactions.
In Fig.~\ref{fig:triadic_dynamos} we present all the triads that lead to linear instabilities and hence to energy transfer,
%%%%%%%%%%%%%% DYNAMO
%%%%%%%%%%%%%%
where the red arrows indicate large-scale dynamo action (i.e., energy is
transferred to small-wavenumber helical modes of the magnetic field), the blue arrows
indicate small-scale dynamo action (i.e., energy is transferred to 
large-wavenumber helical modes of the magnetic field), and the thickness of the arrows
indicates the magnitude of the transfer 
(see Refs.~\cite{Linkmannetal16,Linkmann17b} for details).
{More precisely, the thickness of the arrows is 
qualitative, 
%statement 
reflecting that for any nontrivial
triad geometry the growth rates of the perturbations have a consistent ordering [see, e.g.,
Eq. \eqref{eqapp:growthrates}]}. %or the appendix of Ref.~\cite{Linkmann17b}.}

\begin{figure}[!ht]
\center
\begin{subfigure}{0.12\textwidth}
  \includegraphics[width=\textwidth]{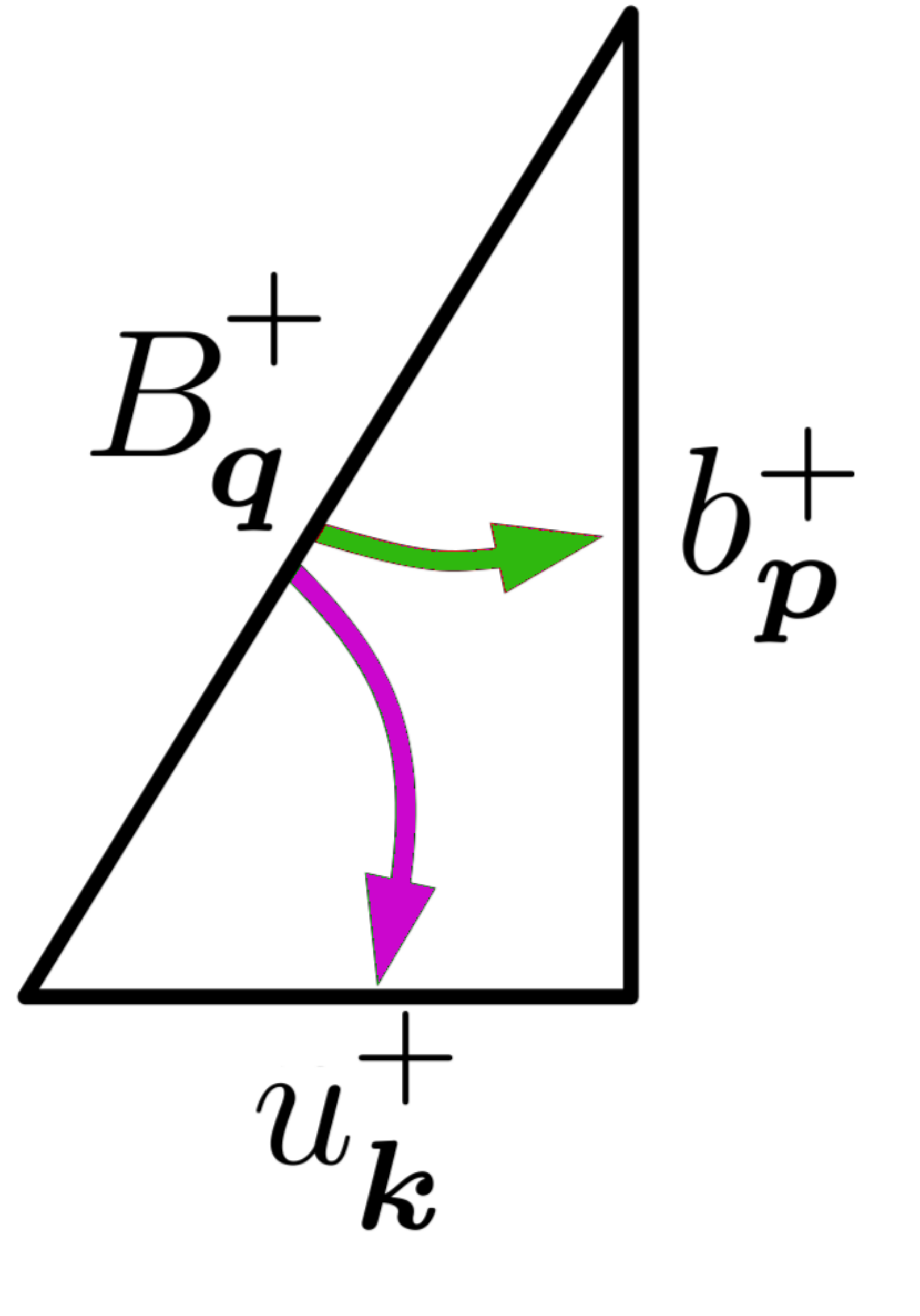}
  \caption{}
  \label{fig:6a}
\end{subfigure}
\begin{subfigure}{0.12\textwidth}
  \includegraphics[width=\textwidth]{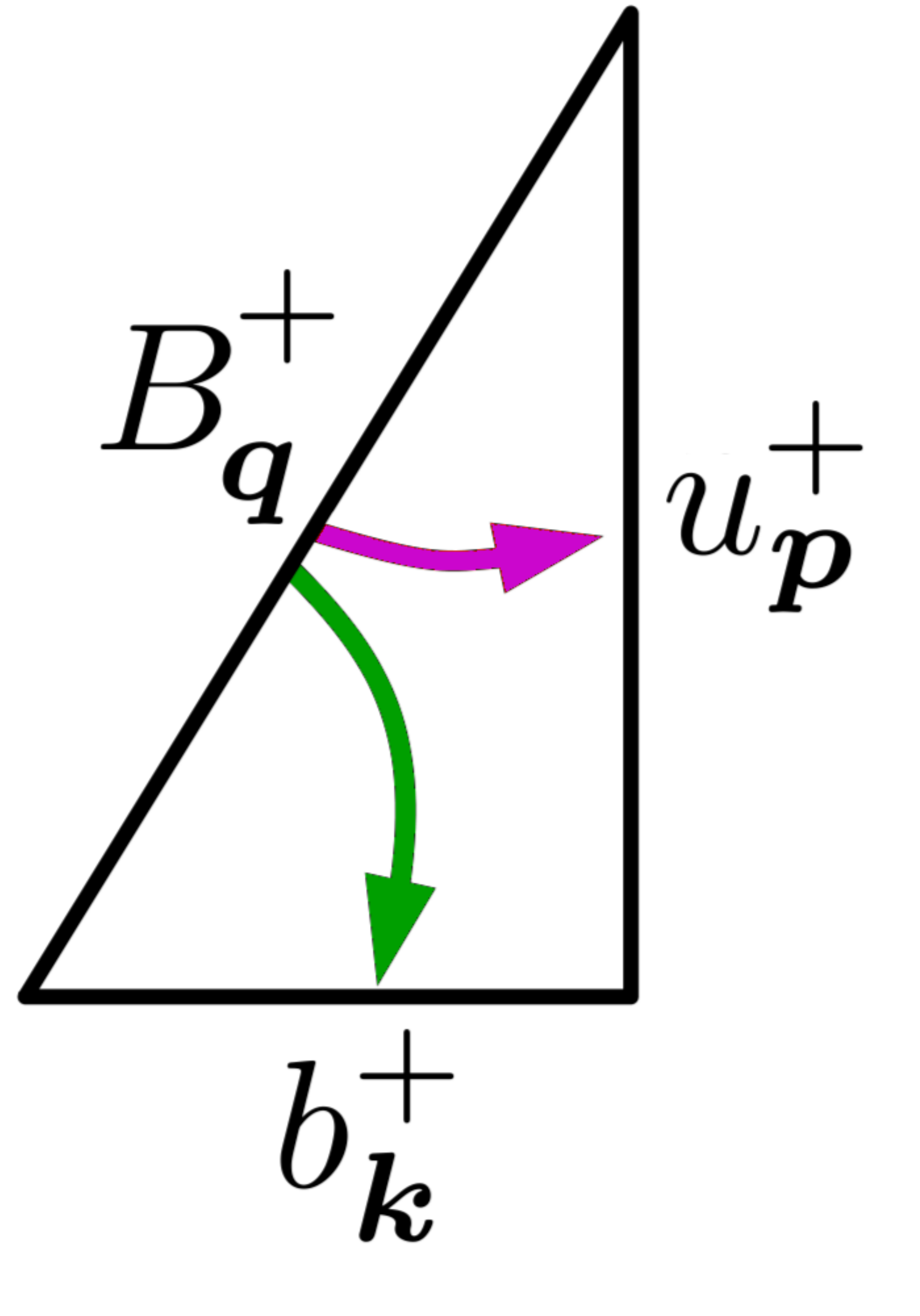}
  \caption{}
  \label{fig:6c}
\end{subfigure}
\begin{subfigure}{0.12\textwidth}
  \includegraphics[width=\textwidth]{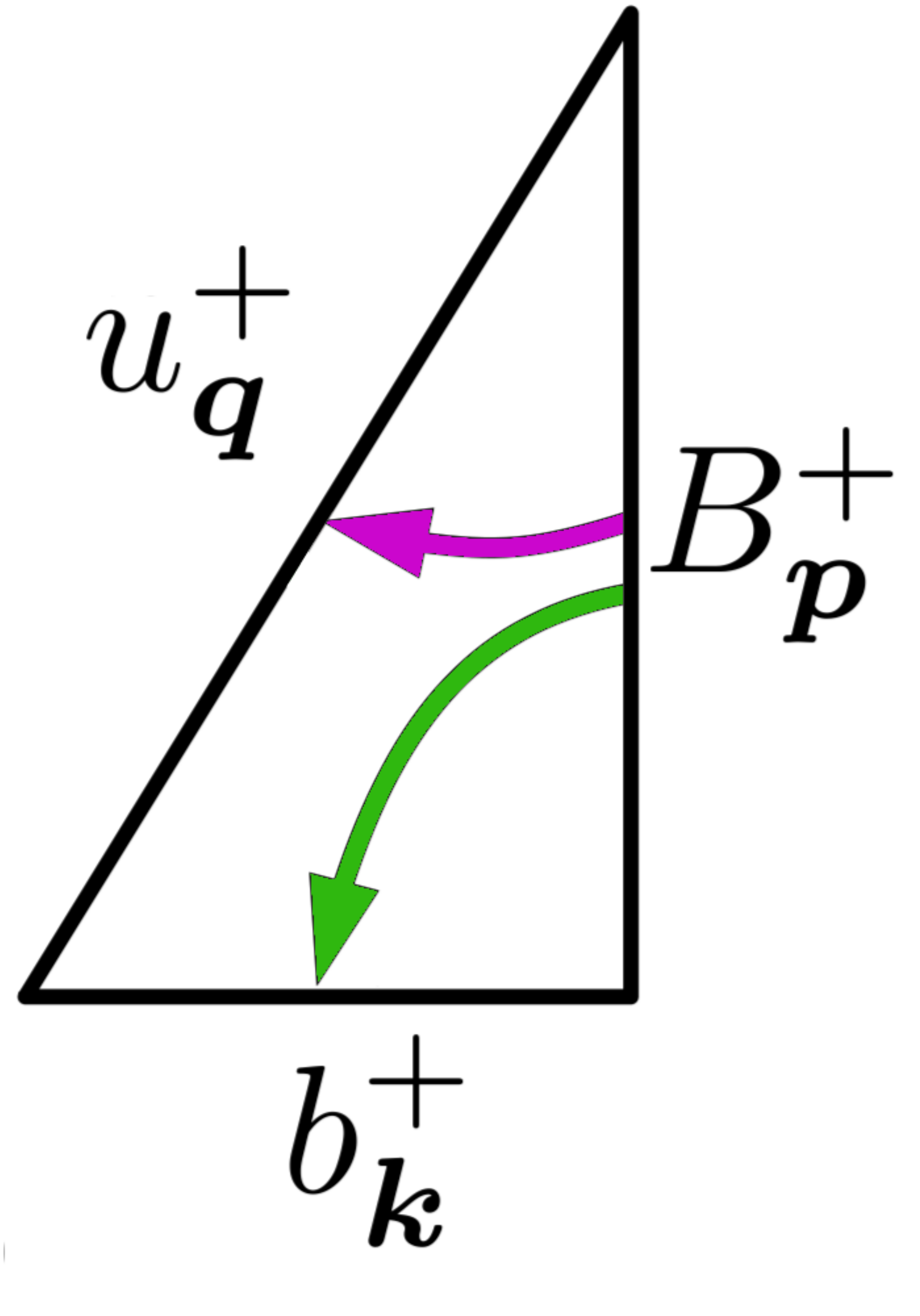}
  \caption{}
  \label{fig:6e}
\end{subfigure}
\\
\begin{subfigure}{0.12\textwidth}
  \includegraphics[width=\textwidth]{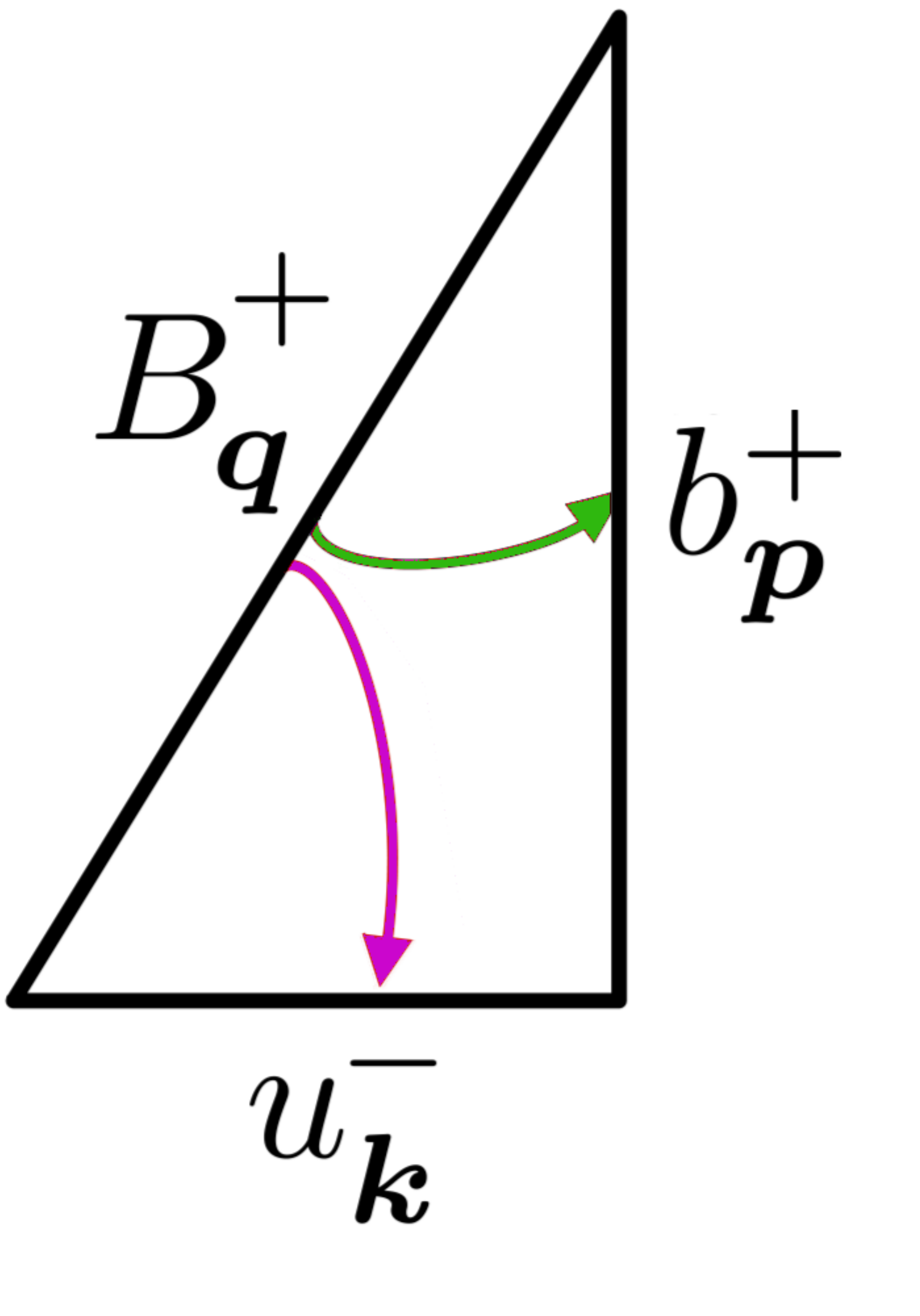}
  \caption{}
  \label{fig:6b}
\end{subfigure}
\begin{subfigure}{0.12\textwidth}
  \includegraphics[width=\textwidth]{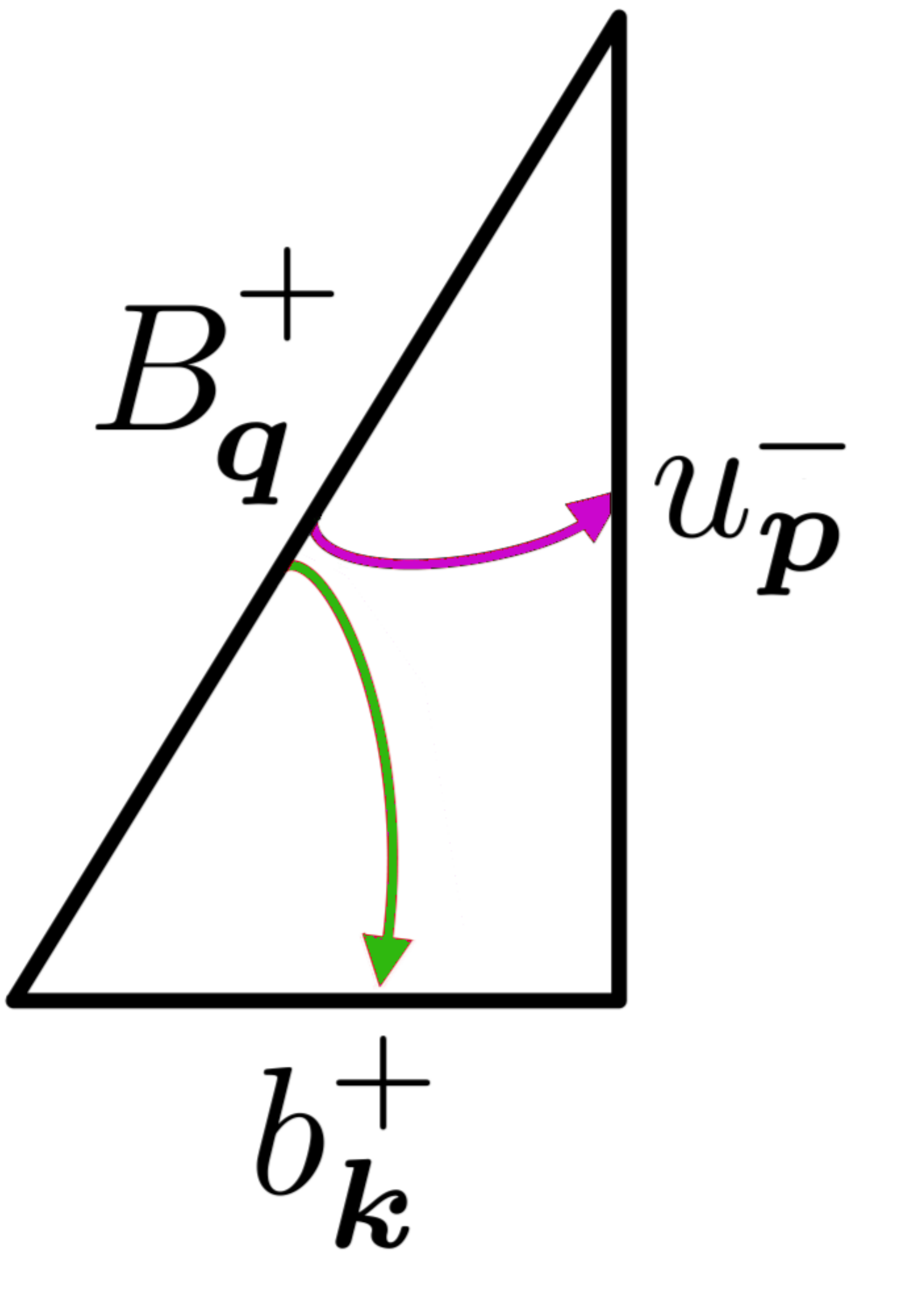}
  \caption{}
  \label{fig:6d}
\end{subfigure}
\begin{subfigure}{0.12\textwidth}
  \includegraphics[width=\textwidth]{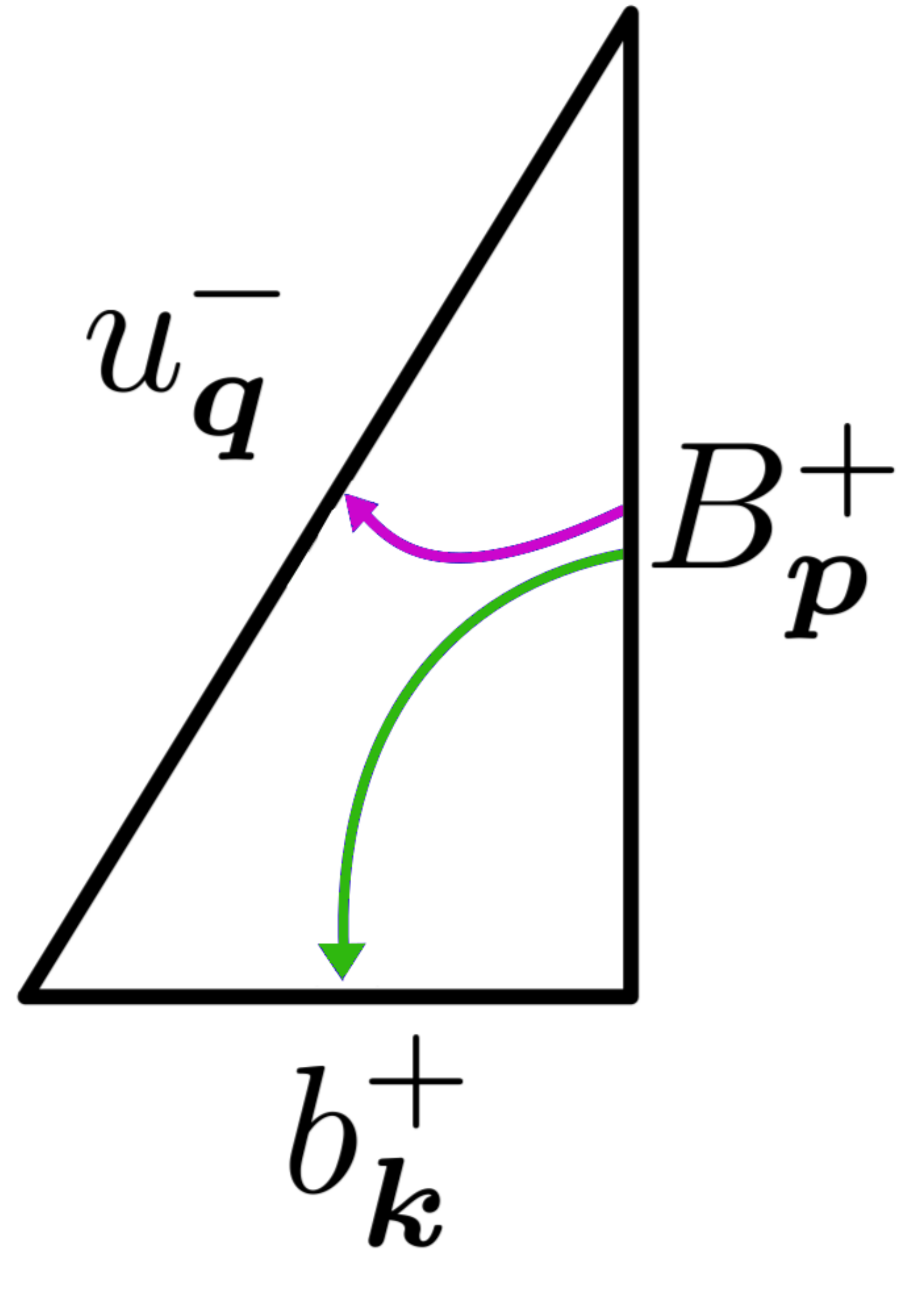}
  \caption{}
  \label{fig:6f}
\end{subfigure}
\caption{
Triadic interactions of helical modes 
leading to energy transfer
when considering a stable
solution for the magnetic field subject to velocity and magnetic perturbations.
Green (light gray) arrows indicate an  
inverse transfer of magnetic helicity and magenta (dark gray) arrows indicate
conversion of magnetic to kinetic energy 
due to the Lorentz force. The thickness of the arrows indicates the magnitude of the transfer.}
\label{fig:triadic_ic_lorentz}
\end{figure}

Similarly, if we consider the linear stability analysis of a steady solution
for the magnetic field subject to velocity and magnetic perturbations, one can
obtain the triadic interactions involving unstable helical modes shown in
Fig.~\ref{fig:triadic_ic_lorentz}, 
%%%%%%%%%%%%%% IC/LORENTZ
%%%%%%%%%%%%%%
where the green arrows indicate magnetic energy transfer, the magenta arrows
indicate the conversion of magnetic to kinetic energy due to the action of the
Lorentz force, and the thickness of the arrows indicates the magnitude of the transfer.
%Following the analysis that was carried out in Ref.~\cite{Linkmann17b} for the triadic 
%interactions corresponding to kinematic dynamo action and the inverse transfer of magnetic helicity, 
%we carry out the linear stability analysis for triad interactions of 
%the Lorentz force in Appendix \ref{app:Lorentz}.
%
As can be seen in Fig.~\ref{fig:triadic_ic_lorentz}, 
triadic interactions of helical modes leading to energy transfer 
occur only if (a) the steady solution of the magnetic field and the 
magnetic perturbation have the same sign
of magnetic helicity, and (b) the characteristic wave number of the magnetic
perturbation is smaller than the characteristic wavenumber of the steady
solution.
In other words, the energy transfer between the magnetic modes %cascade
occurs exclusively from large to small wavenumbers 
and only between magnetic modes with the same 
sign of helicity, %of magnetic helical modes,
which implies an inverse cascade of magnetic helicity \cite{Linkmannetal16}.
As indicated by the thickness of the green arrows, the inverse cascade of magnetic helicity is
stronger if the magnetic and kinetic helicities are of the same sign \cite{Linkmann17b}.
{Following the analysis that was carried out in Ref.~\cite{Linkmann17b} for the triadic 
interactions corresponding to kinematic dynamo action and the inverse transfer of magnetic helicity, 
we carry out an analysis of the relative magnitude of the growth rates corresponding to triad interactions of 
the Lorentz force in the Appendix.}
Our analysis %(see Appendix \ref{app:Lorentz}) 
shows that the conversion of magnetic to kinetic energy 
through the Lorentz force occurs mainly between helical 
modes of the velocity and the magnetic field that have the same sign of helicity,
as indicated by the thick magenta arrows in Fig. \ref{fig:triadic_ic_lorentz}.

\subsection{Direct cascade of magnetic helicity}
In order to understand the presence of the direct cascade of positive magnetic
helicity, observed in our simulations, at the level of triadic interactions, we
aim to identify interactions among helical modes that transfer energy to
helical modes of positive magnetic helicity at large wavenumbers. The only
possible way to get such interactions is via 
the triads in Figs. \ref{fig:5a}, \ref{fig:5c}, and \ref{fig:5b}.
From these three triad interactions, those that
dominate the small scales are shown in Figs.~\ref{fig:5a} and \ref{fig:5c}, where the
unstable helical modes $U_{\vk}^+$ and $U_{\vp}^+$ transfer energy to
$B_{\vq}^+$ via transfers that represent a small scale dynamo. In all these
cases, energy is transferred from a small-wave-number mode to a large-wave-number mode and this
transfer originates from a positively helical unstable mode of the velocity field. As
we have already observed from our simulations, kinetic helicity has a
well-defined spectrum [see Fig.~\ref{fig:rhouk}] and thus positively
helical modes of the velocity field are present 
and can enable such triadic interactions.

The generation of kinetic helicity in these flows can be understood via the
triadic interactions of Fig.~\ref{fig:6a}-\ref{fig:6e}. These
three triads are the dominant interactions in this linear stability analysis
and in all these cases modes of positive magnetic helicity generate modes of
positive kinetic helicity. In other words, the analysis of triadic interactions
suggests that the magnetic helicity generates kinetic helicity 
through triad interactions associated with the Lorentz force.
As discussed in Sec.~\ref{sec:time_evolution}, our flows are %naturally 
dominated by positively
helical modes of the magnetic field because we maintain positive net magnetic
helicity [see Fig.~\ref{fig:rhob}] via the positively helical electromagnetic force $\bm
f_b$.

In summary, since
there is no direct transfer of energy between
modes that have opposite sign of magnetic helicity, %.  So, 
the only way for a
direct cascade of $H_b$ to occur is via the Lorentz force, which generates
modes with positive kinetic helicity. These in turn excite modes with both
positive and negative magnetic helicity with the dominant interactions creating
positively helical magnetic-field modes at small scales (see Figs.
\ref{fig:5a}, \ref{fig:5c}, and \ref{fig:5b}).

\subsection{Negative magnetic helicity at large scales}

Here we discuss %aim to identify 
the triad interactions that lead to the 
generation of modes with negative magnetic helicity at small wave numbers. 
As 
stated above, there is no
direct transfer of energy between modes that have opposite sign of magnetic
helicity.  Therefore, the triad interactions that generate opposite signs of
magnetic helicity between large and small scales have to involve a transfer of
energy from positively helical modes of the velocity field. According to the
results of the stability analysis of the triadic interactions shown in Figs.
\ref{fig:triadic_dynamos} and \ref{fig:triadic_ic_lorentz}, we can conclude
that there are two types of instabilities leading to the occurrence of
large-scale magnetic helicity, namely, triadic instabilities that represent
large-scale dynamo, including STF dynamics [see Figs.~\ref{fig:5c} and
\ref{fig:5e}], and those that can be directly associated with the 
inverse cascade
of positive magnetic helicity [see Figs.~\ref{fig:6c} and \ref{fig:6e}].  
Here we
expect a competition between triad interactions in Figs.~\ref{fig:5c} and
\ref{fig:5e} and those in Figs.~\ref{fig:6c} and \ref{fig:6e}. Based on our
numerical simulations, we observe the persistent generation of negative
magnetic helicity at wave number $k=3$ [see Fig. \ref{fig:rhobk}] despite the
positive mean value of magnetic helicity that is maintained by the positively
helical electromagnetic forcing.  This observation indicates that the triad
interactions of Figs.~\ref{fig:5c} and \ref{fig:5e} dominate over Figs.
\ref{fig:6c} and \ref{fig:6e} at low wavenumbers as the scale separation $k_fL$
increases. 
Thus, the generation of opposite signs of magnetic helicity between
large and small scales can only be partially explained by STF dynamics, which
is essentially what the triad in Fig.  \ref{fig:5c} represents. This is because
the triadic interactions of the helical modes in Fig. \ref{fig:5e}, which
represent a small scale dynamo, can play an important role.
Moreover, the persistence of the triads in Figs. \ref{fig:5c} and \ref{fig:5e} can also
explain why the scales larger that the forcing scale do not become fully
helical and hence the relative helicity $\rho_b(k) < 1$ at wavenumbers $1 < k <
k_f$.

\section{Conclusions} \label{sec:conclusions}
In this paper we investigated the bidirectional cascade of magnetic helicity
$H_b$ in homogeneous MHD turbulence by means of direct numerical simulations
and analysis of triad interactions of helical modes.  In order to be able to
simulate statistically stationary high-$Rm$ flows
with large enough scale separation we considered high-order dissipation terms
acting at small and large scales of the velocity and the magnetic field.  To
maintain positive mean magnetic helicity in our flows we used a helical
electromagnetic forcing, while the velocity field was forced by a nonhelical
forcing.

Our numerical simulations demonstrated the following main results.
(a) Magnetic helicity exhibits a clear bidirectional cascade even in these high-$Rm$ flows.  This bidirectional cascade consists of a dominant inverse cascade
and a residual direct cascade of smaller magnitude, %amplitude
which remains invariant as the scale separation $k_fL$ increases.  
(b) Despite the mean positive magnetic
helicity measured in all our simulations, we consistently observe the
generation of opposite signs of magnetic helicity between large and small
scales.

Our numerical results were interpreted in terms of the triad interactions of the helical Fourier modes.
The residual forward cascade of magnetic helicity may occur due to a nonlinear small-scale dynamo, while the occurrence of large-scale
negative helicity may result from a competition between triadic interactions
associated with dynamo action and those corresponding to the inverse cascade of
magnetic helicity.

The analysis of triadic interactions of helical modes predicts that the kinetic
helicity in our flows is generated at all scales by the unstable modes of
positive magnetic helicity. 
This is deduced from the stability properties of triadic interactions associated 
with the Lorentz force, and the predictions %from \cite{Linkmannetal16,Linkmann17b}. 
are in agreement with our numerical simulations, where kinetic helicity
becomes significant by developing a whole spectrum of 
predominantly positive helicity across the scales.
When the modes of positive kinetic helicity are unstable
they preferentially generate modes of positive magnetic helicity 
at small scales \cite{Linkmannetal16,Linkmann17b}. 
This explains the existence of the residual 
direct cascade of magnetic helicity, 
which is associated with a triad interaction representing 
the Lorentz force and a triad that represents STF dynamics
(i.e. the generation of opposite signs of magnetic helicity between between large and small
scales due to a helical velocity field). 
Moreover, since
triad interactions between modes that have opposite sign of magnetic helicity
are prohibited, the only way for the magnetic helicity to have opposite
signs between large and small scales is via the combined effect of triad
interactions associated with STF dynamics and small-scale dynamo action, which
involve unstable modes of positive kinetic helicity.

A large network of triadic interactions given by the MHD equations can behave
differently from a collection of isolated systems of triads \cite{Moffatt14}.
For example, we cannot comment just from the analysis of triad interactions
whether the direct cascade of magnetic helicity will vanish at the limit $Rm_f
\to \infty$. Therefore, numerical investigations of the effect of different
couplings of helical modes in DNSs like in Refs.~\cite{Biferale12,Alexakis17}
would complement our analysis on the interscale transfers of magnetic helicity.

Preliminary results from numerical simulations with fixed scale separation
$k_fL$ and increasing $Rm_f$ suggest that the direct cascade of magnetic
helicity may be a finite-Reynolds-number effect. This is because the scaling of
the normalized dissipation rate of $H_b$ at small scales is
$\veps^+_{H_b}/\veps_{H_b} \propto Rm_f^{-0.22}$. This weak power law implies
that the direct cascade of magnetic helicity will persist as $Rm_f$ increases,
but it will eventually vanish in the limit of $Rm_f \rightarrow \infty$. This claim is
supported by deriving upper bounds on the absolute value of the dissipation
rate of $H_b$ at small scales. 
However, further investigation is required in
order to understand better the small scale dynamics of magnetic helicity. In
principle, this could be achieved through a series of high-resolution DNSs of
helical MHD turbulence covering a large range of magnetic Reynolds numbers from
which one could extrapolate towards the high-Reynolds-number asymptotic limit.
However, such a series of simulations will be computationally challenging as
sufficient scale separation is required between the forcing scale and the box
size, while the use of the Laplacian operator requires a large number of
collocation points in order to achieve adequate small scale resolution.

{Magnetic helicity is considered to play a fundamental role in the dynamo
action of planetary, stellar, and galactic magnetic fields. The time scale and
the level of the saturation of the magnetic field in models of the
solar activity cycle depend on assumptions about the diffusion of magnetic
helicity by small-scale velocity fluctuations \cite{kleeorinetal03}. Our
results improve our understanding of the magnetic helicity dynamics on the
formation and evolution of large-scale magnetic fields in astrophysical
objects, which is one of the challenging problems in modern astrophysical fluid
dynamics.  }

\textit{Note added.}
Recently, it has come to our attention that similar conclusions concerning the
behavior of the direct cascade of magnetic helicity in the limit of infinite 
$Rm$ have been reached \cite{Aluie17}.

\section*{Acknowledgements}
We thank Luca Biferale and Ganapati Sahoo for helpful conversations.
This work has made use of the resources provided by ARCHER, 
made available through the Edinburgh Compute and
Data Facility (ECDF).  V.~D. acknowledges
support from the Royal Society and the British Academy of Sciences (Newton
International Fellowship, Grant No.~NF140631).  The research leading to these results has
received funding from the European Union's Seventh Framework Programme
under grant agreement No. 339032.  

\appendix
\section{Triadic analysis of the Lorentz force} \label{app:Lorentz}
Following the work of Refs.~\cite{Linkmannetal16,Linkmann17b}, we present % carry out 
here the linear stability analysis of the triadic interactions that correspond 
to the Lorentz force as an example. This analysis allows us to derive the additional results concerning 
the magnitude of the magnetic-to-kinetic energy transfers depicted in 
Fig.~\ref{fig:triadic_ic_lorentz}. We consider a positively helical magnetic steady solution 
$(\Bpp \ne 0, \Bpm =\Upm =\Upp= 0)$ of Eqs.~\eqref{eq:basic-triads} 
at wave vector $\vp_0 = \vp$ 
subject to magnetic and velocity field perturbations %at wavevectors 
$\tbkp$, $\tbkm$, $\tuqp$, and $\tuqm$, \\
\begin{minipage}[l]{.5\textwidth}
  \centering
\includegraphics[scale=0.1]{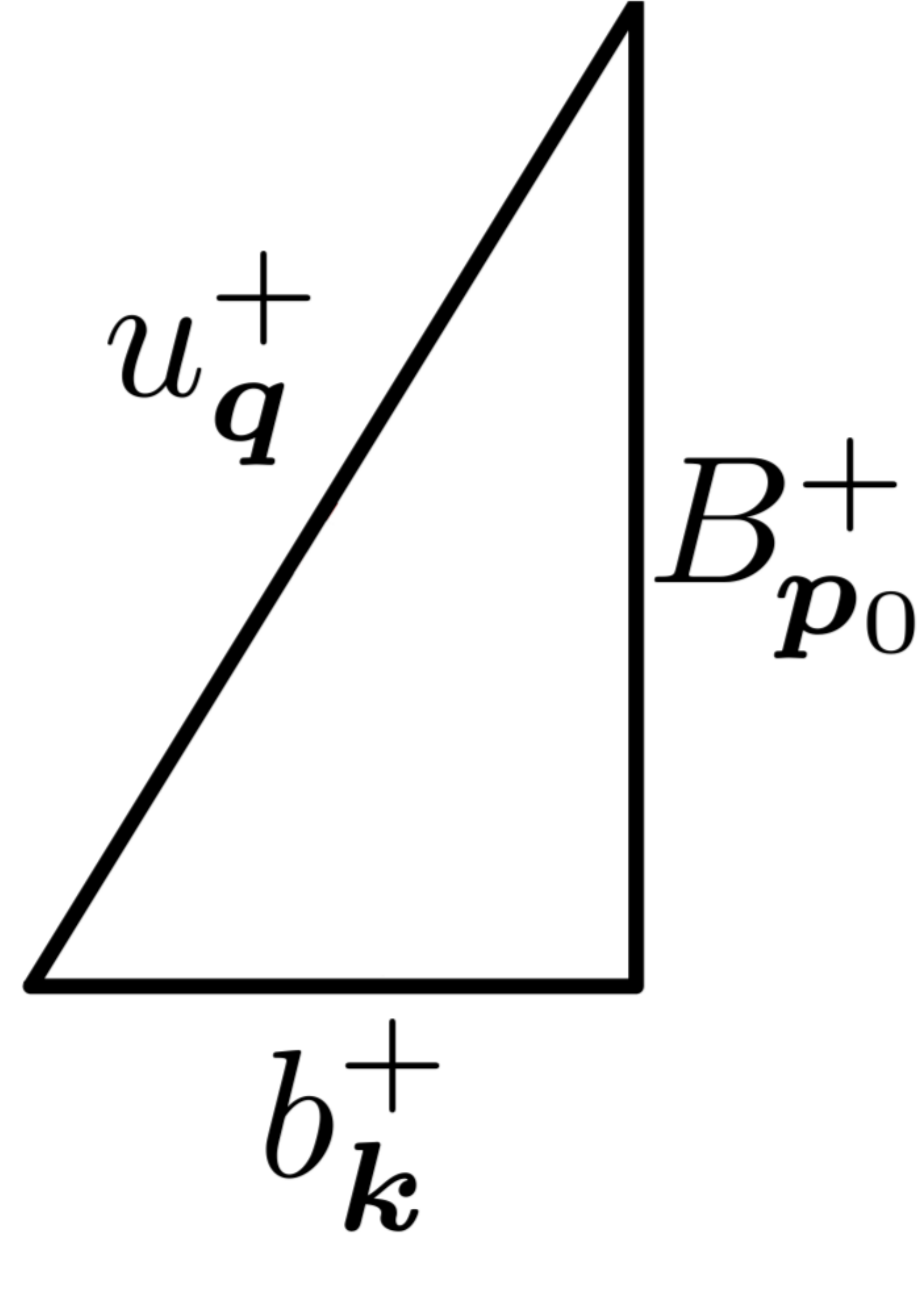} \\
\includegraphics[scale=0.1]{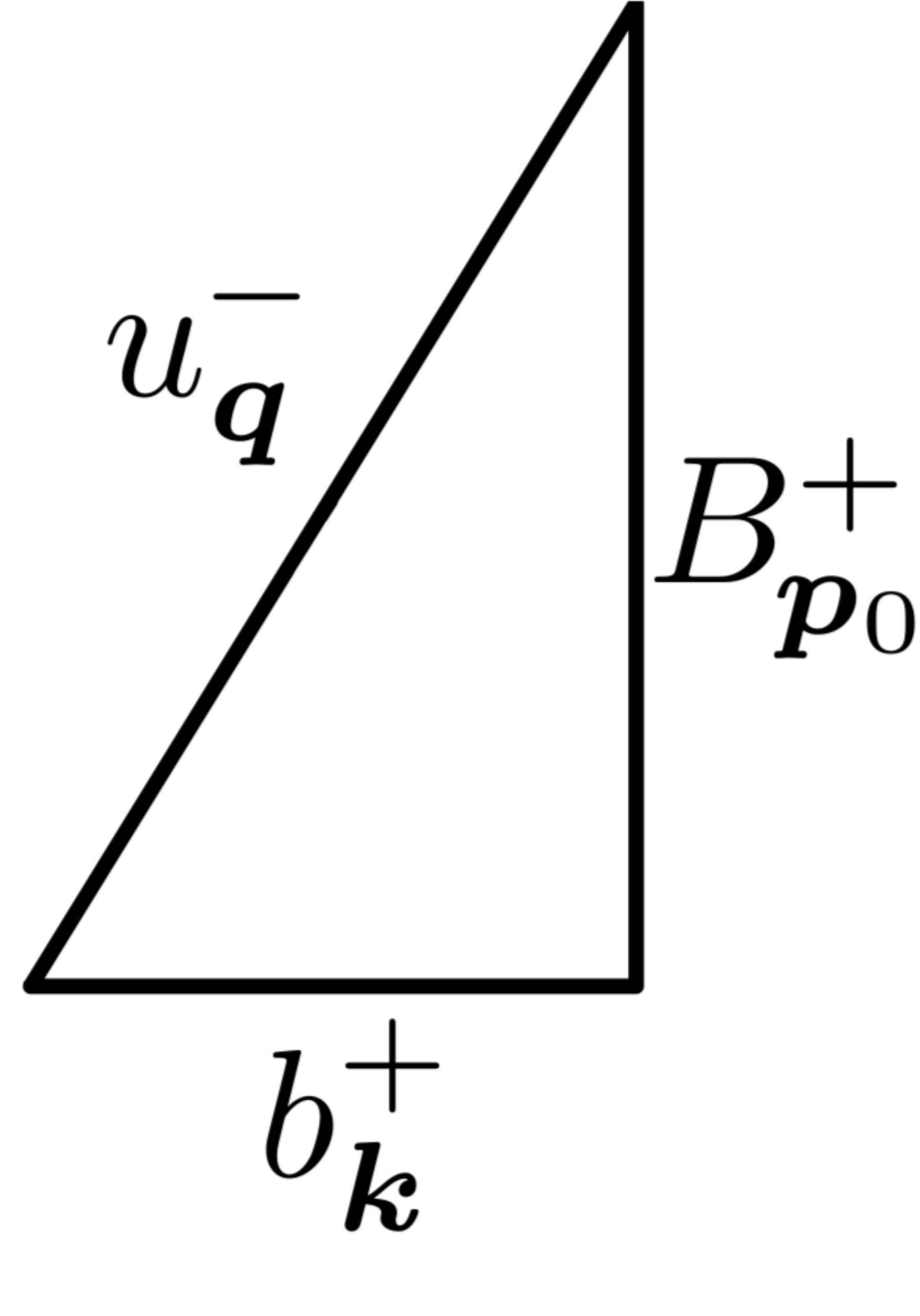} \\
\includegraphics[scale=0.1]{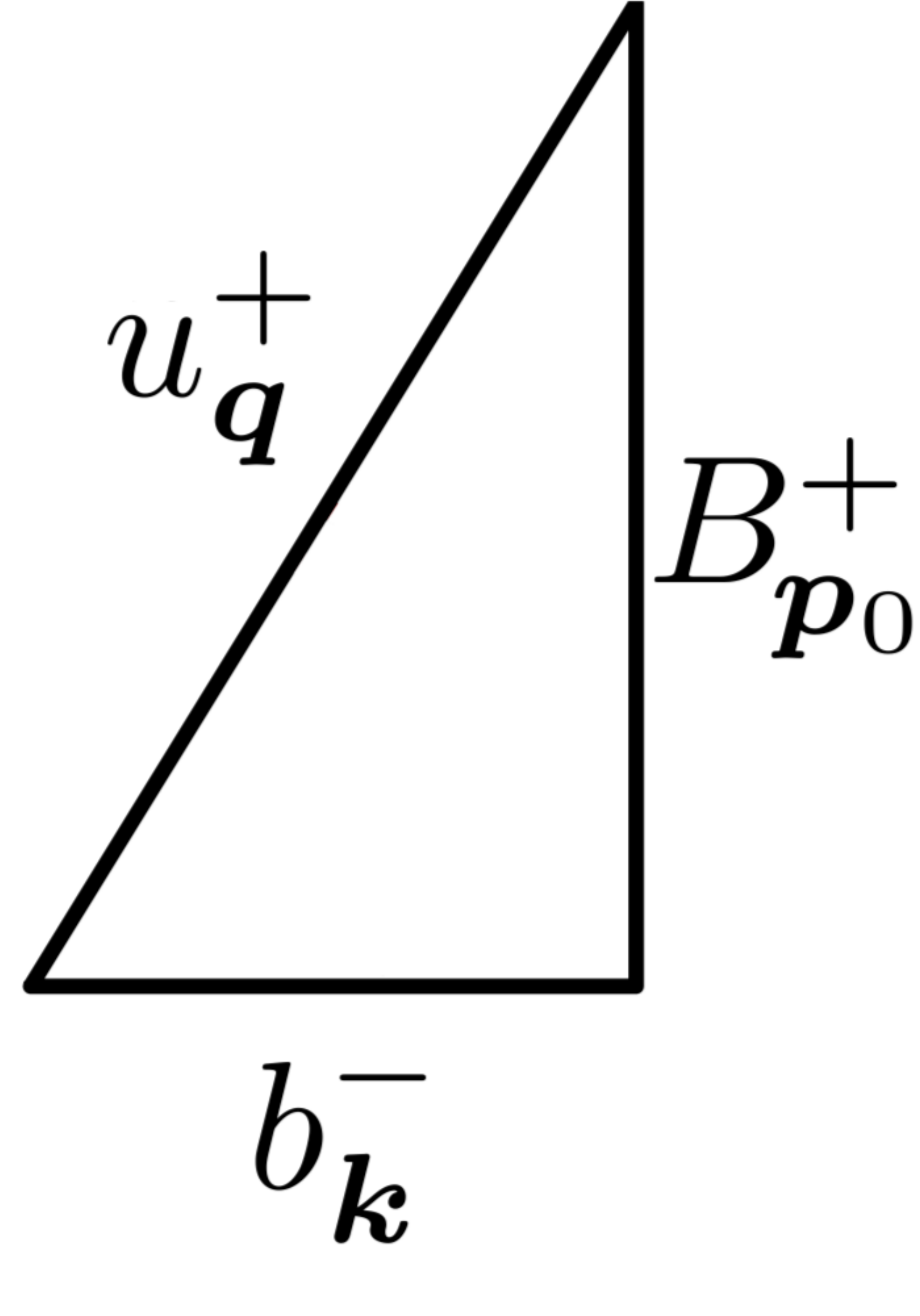} \\
\includegraphics[scale=0.1]{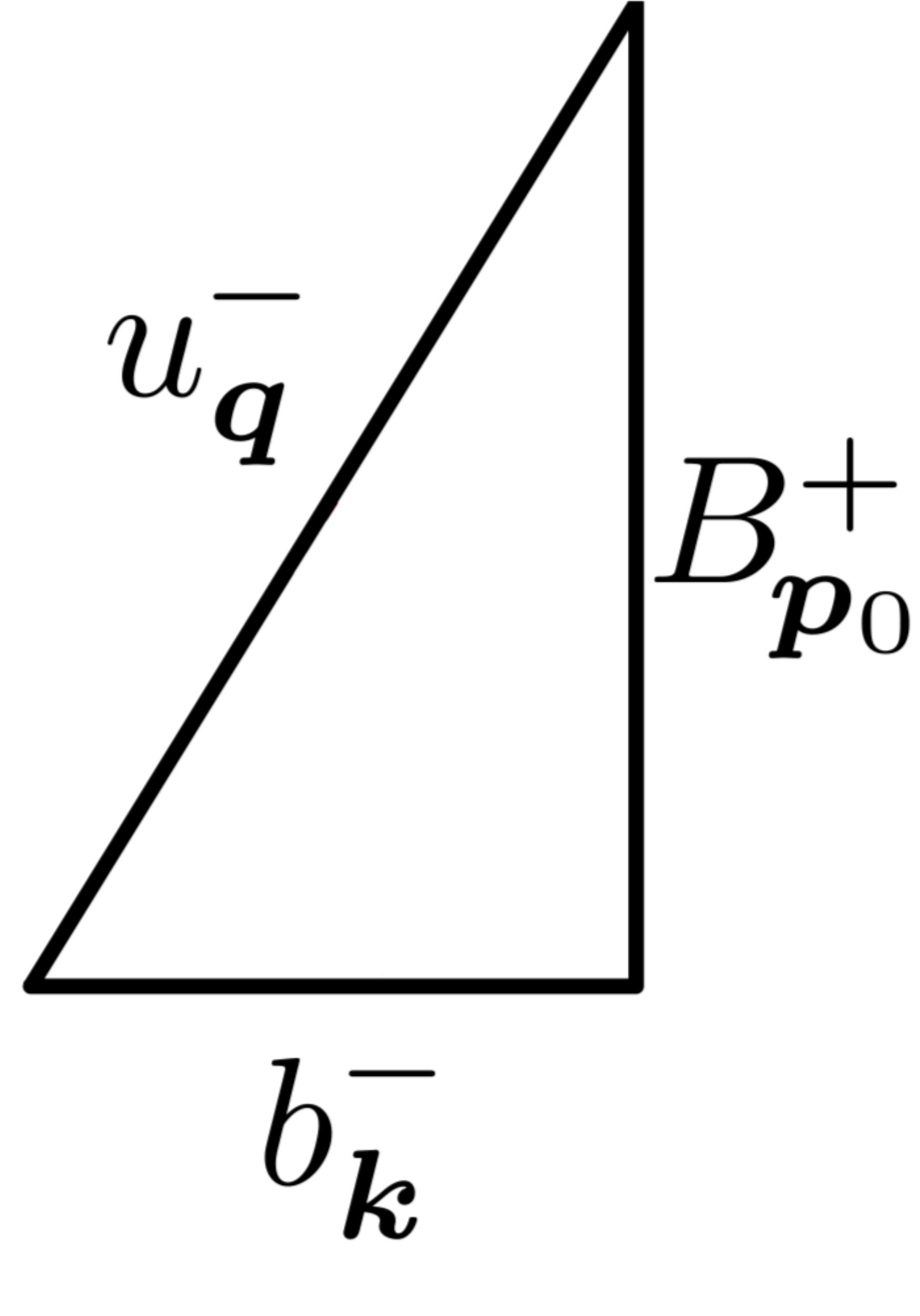}
\end{minipage}
\hspace{-1cm}
\vspace{-0.8cm}
\begin{minipage}[l]{.5\textwidth}
\vspace{-0.7cm}
\begin{align}
\label{eq:triad1}
& %LF1 \quad
\begin{cases}
\dt \tbkp  =  k \:  g_{+++} \Bpp \tuqp  \ , \\ 
\dt \tuqp  = - g_{+++} \:(p_0-k)\: \Bpp \tbkp \ , 
\end{cases} \qquad  \qquad\\
\nonumber \\
\nonumber \\
\nonumber \\
\label{eq:triad2}
& %LF2 \quad
\begin{cases}
\dt \tbkp  =  k \:  g_{++-} \Bpp \tuqm  \ , \\ 
\dt \tuqm  = - g_{++-} \:(p_0-k)\: \Bpp \tbkp \ , 
\end{cases} \qquad  \qquad\\
\nonumber \\
\nonumber \\
\nonumber \\
\label{eq:triad3}
& %LF3 \quad
\begin{cases}
\dt \tbkm  =  -k \:  g_{-++} \Bpp \tuqp  \ , \\ 
\dt \tuqp  =  g_{-++} \:(p_0+k)\: \Bpp \tbkm \ , 
\end{cases} \qquad  \qquad\\
\nonumber \\
\nonumber \\
\nonumber \\
\label{eq:triad4}
& %LF4 \quad
\begin{cases}
\dt \tbkm  =  -k \:  g_{-+-} \Bpp \tuqm \ , \\ 
\dt \tuqm  = g_{-+-} \:(p_0+k)\: \Bpp \tbkm \ , 
\end{cases} \qquad \qquad
\nonumber \\
\end{align}
\end{minipage}\\

\vspace{1cm}
%\clearpage
\noindent
where each set of equations represents a particular helical combination of 
velocity- and magnetic-field modes from the 
system of equations \eqref{eq:basic-triads} for a particular choice of 
helicities. 
The procedure is conceptually similar to a linear stability analysis 
of rigid body rotation \cite{Waleffe92}, that is,  
we differentiate Eqs.~\eqref{eq:triad1}-\eqref{eq:triad4} further in time and
replace any occurrence of a first-order time derivative 
by Eqs.~\eqref{eq:basic-triads}. 
In order to isolate the dynamical effects of the Lorentz force on the flow
we focus on the evolution of the velocity perturbations
\begin{align}
\label{eq:uplusLF1} 
%LF1 \qquad 
\partial_t^2 \tuqp &= (p_0-k)k |g_{+++}|^2 \, |\Bpp|^2 \tuqp \ , \\ 
\label{eq:uplusLF2} 
%LF2 
\qquad \partial_t^2 \tuqm &=  (p_0-k)k |g_{++-}|^2 \, |\Bpp|^2  \tuqm \ , \\ 
\label{eq:uminusLF1} 
%LF3 
\qquad \partial_t^2 \tuqp &= -(p_0+k)k |g_{-++}|^2 \, |\Bpp|^2 \tuqp \ , \\ 
\label{eq:uminusLF2} 
%LF4 
\qquad \partial_t^2 \tuqm &= -(p_0+k)k |g_{-+-}|^2 \, |\Bpp|^2 \tuqm \ , 
\end{align}
and we observe that Eqs.~\eqref{eq:uminusLF1} and \eqref{eq:uminusLF2} cannot
have exponentially growing solutions as the term $-(p_0+k)k$ is always negative. 
Concerning Eqs.~\eqref{eq:uplusLF1} and \eqref{eq:uplusLF2}, the wavenumber ordering $k<p_0<q$ results in 
$(p_0-k)k > 0$, hence these two equations admit exponentially growing solutions. 
In other words, linear instabilities occur only in Eqs.~\eqref{eq:uplusLF1} and \eqref{eq:uplusLF2}. 

From Eqs.~\eqref{eq:triad1} and \eqref{eq:triad2} we
observe that the coefficient $g_{+++}$ corresponds to $\tuqp$ and $\tbkp$ coupling to
$\Bpp$. Therefore, the term $(p_0-k)k |g_{+++}|^2 \, |\Bpp|^2$
in Eq.~\eqref{eq:uplusLF1} describes the coupling of 
magnetic-field modes with the same signs of helicity, i.e., $\tbkp$ and $\Bpp$ to $\tuqp$. 
Similarly,
the term $-(p_0+k)k |g_{-++}|^2 \, |\Bpp|^2$ in
Eq.~\eqref{eq:uminusLF1} describes the coupling of magnetic-field
modes with opposite signs of helicity to $\tuqp$, because $g_{-++}$ couples $\tuqp$ to $\Bpp$ and $\tbkm$. 
Equations.~\eqref{eq:uplusLF2} and \eqref{eq:uminusLF2} are analyzed analogously. 
Hence linear instabilities only occur in triadic interactions 
where the unstable magnetic mode and the magnetic perturbation have the same sign of
helicity \cite{Linkmannetal16}.  
In terms of the instability assumption this result implies that 
the Lorentz force can only convert magnetic to kinetic energy
if the interaction proceeds by triads involving magnetic-field modes with 
the same sign of helicity \cite{Linkmannetal16}.

The energy transfer due the linear instability governed by Eq.~\eqref{eq:uplusLF1}
corresponds to the magenta arrow in Fig.~\ref{fig:6c}, while the linear instability 
described by Eq.~\eqref{eq:uminusLF1}
corresponds to the magenta arrow in Fig.~\ref{fig:6f}. The remaining energy transfers depicted in 
Fig.~\ref{fig:triadic_ic_lorentz} can be derived analogously by changing the characteristic wavenumber of the
steady magnetic-field mode. 
The magnitude of the transfers indicated by the thickness of the arrows in Fig.~\ref{fig:triadic_ic_lorentz}
is determined by a comparison of the growth rate of the mechanical perturbations
\beq
\label{eqapp:growthrates}
\frac{(p_0-k)k |g_{-++}|^2 \, |\Bpp|^2}{(p_0-k)k |g_{+++}|^2 \, |\Bpp|^2} = \frac{|g_{-++}|^2}{|g_{+++}|^2} \leqslant 1 \ ,
\eeq
since the magnitude of the geometric factors can be written as $|g_{s_ks_ps_q}|
= |s_kk +s_pp +s_qq|[2(k^2p^2+p^2q^2+q^2k^2)-k^4-p^4-q^4]^{1/2}$
\cite{Waleffe92,Linkmann17b}. Hence the energy transfer from a positively
helical magnetic field into a positively helical flow is stronger than that
occurring from a positively helical magnetic field into a negatively helical
flow. This result is indicated 
by a thicker magenta arrow in Fig.~\ref{fig:6e} 
than in Fig.~\ref{fig:6f}. 
The relative magnitudes of the remaining energy transfers connected with dynamo
action as shown in Figs.~\ref{fig:triadic_dynamos} and the inverse cascade of
magnetic helicity shown by the green arrows in Fig.~\ref{fig:triadic_ic_lorentz}
have been derived by the same method in Ref.~\cite{Linkmann17b}.

\bibliographystyle{unsrt}
\bibliography{allrefs}

\end{document}